\documentclass[aps,prc,showpacs,amsfonts,twocolumn,floatfix,superscriptaddress]{revtex4-1}
\usepackage{graphicx,epsfig,amssymb,color,amsmath}
\usepackage{natbib}
\usepackage{ulem}
\usepackage{braket}
\usepackage{cases}
\usepackage{color}
\usepackage{mathtools}
\usepackage{caption}
\usepackage{subcaption}

\newcommand{\D}{\text{d}}
\DeclareMathAlphabet{\mathpzc}{OT1}{pzc}{m}{it}

\begin{document}

\title{Description of nuclear systems with a self-consistent configuration-mixing approach. \\ I: Theory, algorithm, and application to the $^{12}$C test nucleus}

\author{C. Robin}
\altaffiliation{Present address: Department of Physics, Western Michigan University, Kalamazoo, MI 49008-5252, USA}
\email{caroline.robin@wmich.edu}
\affiliation{CEA, DAM, DIF, F-91297 Arpajon, France}

\author{N. Pillet}
\email{nathalie.pillet@cea.fr}
\affiliation{CEA, DAM, DIF, F-91297 Arpajon, France}

\author{D. \surname{Pe\~na Arteaga}}
\altaffiliation{Present address: Institut d'Astronomie et d'Astrophysique, Universit\'e Libre de Bruxelles, CP-226, B-1050 Brussels, Belgium}
\affiliation{CEA, DAM, DIF, F-91297 Arpajon, France}

\author{J.-F. Berger}
\affiliation{CEA, DAM, DIF, F-91297 Arpajon, France}


\begin{abstract}
\begin{description}
\item[Background] Although self-consistent multi-configuration methods have been used for decades to address the description of atomic and 
molecular many-body systems, only a few trials have been made in the context of nuclear structure.
\item[Purpose] This work aims at the development of such an approach to describe in a unified way various types of correlations in nuclei, in a self-consistent manner where the mean-field is improved as correlations are introduced. The goal is to reconcile the usually set apart Shell-Model and Self-Consistent Mean-Field methods.
\item[Method] This approach is referred as "variational multiparticle-multihole configuration mixing method". It is based on a double variational principle which yields a set of two coupled equations that determine at the same time the expansion coefficients of the many-body wave function and the single particle states. The solution of this problem is obtained by building a doubly iterative numerical algorithm.
\item[Results] The formalism is derived and discussed in a general context, starting from a three-body Hamiltonian. Links to existing many-body techniques such as the formalism of Green's functions are established. First applications are done using the two-body D1S Gogny effective force. The numerical procedure is tested on the $^{12}$C nucleus in order to study the convergence features of the algorithm in different contexts. Ground state properties as well as single-particle quantities are analyzed, and the description of the first $2^+$ state is examined.
\item[Conclusions] The self-consistent multiparticle-multihole configuration mixing method is fully applied for the first time to the description of a test nucleus. This study allows to validate our numerical algorithm and leads to encouraging results. In order to test the method further, we will realize in the second article of this series, a systematic description of more nuclei and observables obtained by applying the newly developed numerical procedure with the same Gogny force. As raised in the present work, applications of the variational multiparticle-multihole configuration mixing method will however ultimately require the use of an extended and more constrained Gogny force.
\end{description}
\end{abstract}


\maketitle


\section{Introduction}
In the last decades, important progress has been achieved toward a theoretical description of nuclear systems. Extensions of existing many-body 
techniques as well as developments of novel approaches have emerged. In particular, great effort is now devoted to reach an \textit{ab-initio} description of nuclei \cite{lattEFT,NCSM,GFMC,FY,imsrg,cc,scgf}. However, due to the high numerical costs, the most exact approaches are still mainly applicable to light nuclei. To tackle the rest of the nuclear chart, the more phenomenological Self-Consistent Mean-Field (SCMF) method \cite{ring} and Shell-Model (SM) \cite{SM} remain among the most used and powerful approaches. The SCMF method and its extensions are based on the determination of self-consistent orbitals, considering the wave function of the nucleus as a particle-independent state. The idea is to enrich the one-body picture in order to minimize the effect of the residual interaction. Account for missing correlations is usually achieved in a second step via symmetry-breaking and restoration techniques \cite{BlaizotGogny,GognyPadjen,BlaizotBerger,Peru1,Donno1,Donno2,Peru2,Peru3,Delaroche,Gambacurta1,Gambacurta2,Gambacurta3,Robledo,Rodriguez,RodriguezEgido, Valor, Niksic,Bender1,Yao,Bender2,Bally}. The SM on the other hand usually uses a frozen oscillator basis to build a wave function that explicitly preserves symmetries. The active nucleons determining the properties of the system are restricted to a valence space and interact through a renormalized interaction.
\\
\\
This work is taking part in the development of an alternative approach to the nuclear many-body problem, namely the "variational multiparticle-multihole (MPMH) configuration mixing approach", which aims to take advantage of both previous types of methods. The nuclear state is expanded on a set of configurations, and both the mixing coefficients and the single-particle orbitals used to build the Slater determinants, are determined at the same time {\it via} a variational principle. This procedure allows a unified treatment of long-range correlations, preserving at best the fundamental symmetries of the nuclear Hamiltonian. Full self-consistency is obtained since the mean-field and the orbitals evolve according to the correlation content of the nucleus. The MPMH configuration mixing method is in fact the adaptation to nuclear systems of techniques already widely employed in the context of atomic physics and quantum chemistry. In these domains, this type of approach is known as Multi-Configuration Hartree-Fock (MCHF) \cite{MCHF1,MCHF2} or Multi-Configuration Self-Consistent Field (MCSCF) method \cite{MCSCF1, MCSCF2} and leads to very successful results. In nuclear physics, the lack of knowledge of the nuclear force as well as the presence of two types of particles represent additional difficulties. 
\\
\\
Pioneering work using a MCHF-type approach in the context of nuclear physics has been done a few decades ago \cite{Faessler1,Faessler1b}. These first studies, restricted to simple analytical models, were followed by realistic applications to the description of a few nuclei of the $sd$-shell in the intrinsic frame \cite{Satpathy,Faessler2,HoKim,Faessler3,Faessler5}. Due to the limited numerical means at the time, these analyses were however restricted to a small number of configurations. The construction of a generalized single-particle basis in the context of the Random-Phase Approximation (RPA) theory was also mentionned in \cite{VariaRPA} and applied analytically to the Lipkin model.
\\
Recent applications of the MPMH configuration mixing method have been realized using the D1S Gogny interaction \cite{D1S}. However these works did not apply the full self-consistent formalism. For example, analyses of the spectroscopy of $sd$-shell nuclei \cite{Pillet2,LeBloas} were performed using frozen Hartree-Fock orbitals. An earlier work, which presented the complete formalism in the case of an effective density-dependent nuclear interaction, applied the MPMH configuration mixing approach to the description of pairing correlations in the ground states of Sn isotopes making drastic approximations in the equation determining the single-particle states \cite{Pillet1}. A prior study with a similar approximation was also performed using the Skyrme SIII interaction for the mean-field and a residual contact interaction, to describe $K$-isomers in the $^{178}$Hf mass region \cite{Pillet0}.
\\
\\
In the present work we apply for the first time the complete formalism of the MPMH configuration mixing approach in a realistic case \cite{Robin}. In section \ref{section1} we expose and analyze the formalism of the method as a many-body theory. A focus on the understanding of the equation determining the orbitals is made and a connection to the formalism of Green's functions is established. For a precise analysis, and in the view of eventual future applications with different interactions, we consider in this formal part a general three-body Hamiltonian. In section \ref{section2} we expose in detail the numerical algorithm that is used to solve the set of coupled equations. As in this work the numerical calculations are performed using the phenomenological density-dependent D1S Gogny force, a focus is made on the rewriting of the equations and their interpretation according to section \ref{section1}. In section \ref{Results} the solution procedure is applied to the description of the ground state of a test nucleus which we chose to be $^{12}$C. Two different truncation schemes are employed to select the configurations included in the wave function. We compare the convergence features of both schemes, as well as results concerning the properties of the ground state and the single-particle spectrum. Particular attention is paid to the effect of the orbital optimization. Finally, the first $2^+$ excited state is investigated, and excitation energies as well as transition probabilities B(E2) are calculated. In section \ref{conclu} we give conclusions and perspectives to this work.


\section{Formalism}
\label{section1}
In this section we derive the formalism of the MPMH configuration mixing approach from a more general point of view than the one adopted in \cite{Pillet1}. This will allow to emphasize the connection with well-known many-body techniques, such as the Green's functions formalism. To start the discussion, we consider a general three-body nuclear Hamiltonian
\begin{eqnarray}
\hat{H} &=& \hat{K}+\hat{V}^{2N}+\hat{V}^{3N} \nonumber \\
        &=& \sum_{ij} K_{ij} a^{\dagger}_i a_j + \frac{1}{4} \sum_{ijkl} \braket{ij|\widetilde{V^{2N}}|kl}a^{\dagger}_i a^{\dagger}_j a_l a_k \nonumber \\
        && + \frac{1}{36} \sum_{ijklmn} \braket{ijk|\widetilde{V^{3N}}|lmn}a^{\dagger}_i a^{\dagger}_j a^{\dagger}_k a_n a_m a_l \; , 
\label{e:H_3N}               
\end{eqnarray} 
where $\hat{K}$ is the kinetic energy operator, and $\widetilde{V^{2N}}$ and $\widetilde{V^{3N}}$ are the antisymmetrized two- and three-body nuclear interactions respectively. The two-body term $\hat{V}^{2N}$ also includes the Coulomb force.
\\
The "exact" nuclear states $\ket{\Psi_{ex}}$ solutions of the Schroedinger equation 
\begin{equation}
\hat H\ket{\Psi_{ex}}=E\ket{\Psi_{ex}} \; ,
\end{equation}
can be expressed as a superposition of many-nucleon configurations $\ket{\phi_\alpha}$ built on a (complete) single-particle basis $\{ i \}$:
\begin{equation}
\ket{\Psi_{ex}} = \sum_\alpha A_\alpha \ket{\phi_\alpha} \; ,
\label{e:wf}
\end{equation}
with
\begin{equation}
\ket{\phi_\alpha} = \prod_{i \in \alpha} a^\dagger_i \ket{0} \; ,
\end{equation}
where $\ket{0}$ denotes the true particle vacuum.
\\
Each Slater determinant $\ket{\phi_\alpha}$ can also be expressed as a multiple particle-hole excitation of a reference state $\ket{\phi_0}$ associated with a given mean-field:
\begin{equation}
\ket{\phi_\alpha} = \prod_{i}^{M_\alpha} \left( a^\dagger_{i_p} a_{i_h} \right) \ket{\phi_0} \; ,
\label{e:config}
\end{equation}
with
\begin{equation}
\ket{\phi_0} = \prod_{i=1}^A a^\dagger_i \ket{0} \; .
\end{equation}
In Eq.~(\ref{e:config}), the indices $h$ (respectively $p$) stand for "hole'' (respectively "particle'') and denote occupied (respectively unoccupied) orbitals in $|\phi_0\rangle$. $M_{\alpha}$ is called the excitation order of the configuration $|\phi_{\alpha}\rangle$ and corresponds to the number of \textit{p-h} excitations applied to $|\phi_0\rangle$ in order to obtain $ |\phi_{\alpha}\rangle$. The reference state $|\phi_0\rangle$ characterized by $M_{\alpha}=0$ is included in expansion (\ref{e:wf}). Finally, any A-nucleon configuration $\ket{\phi_\alpha}$ is a direct product of proton ($\pi$) and neutron ($\nu$) Slater determinants so that \\ \\
\begin{eqnarray}
\ket{\phi_\alpha} &=& \ket{\phi_{\alpha_\pi}} \otimes \ket{\phi_{\alpha_\nu}} \nonumber \\
                  &=& \prod_{i}^{M_{\alpha_\pi}} \left( a^\dagger_{i_{p_\pi}} a_{i_{h_\pi}} \right) \ket{\phi_{0_\pi}} \otimes \prod_{j}^{M_{\alpha_\nu}} \left( a^\dagger_{j_{p_\nu}} a_{j_{h_\nu}} \right) \ket{\phi_{0_\nu}} \; , \nonumber \\
\end{eqnarray}
and
\begin{eqnarray}
\ket{\phi_0} &=& \ket{\phi_{0_\pi}} \otimes \ket{\phi_{0_\nu}} \nonumber \\
             &=& \prod_{i=1}^Z a^\dagger_{i_\pi} \ket{0} \otimes \prod_{j=1}^N a^\dagger_{j_\nu} \ket{0} \; .
\end{eqnarray}

\begin{figure}[h]
\begin{center}\includegraphics[width=0.6\columnwidth]{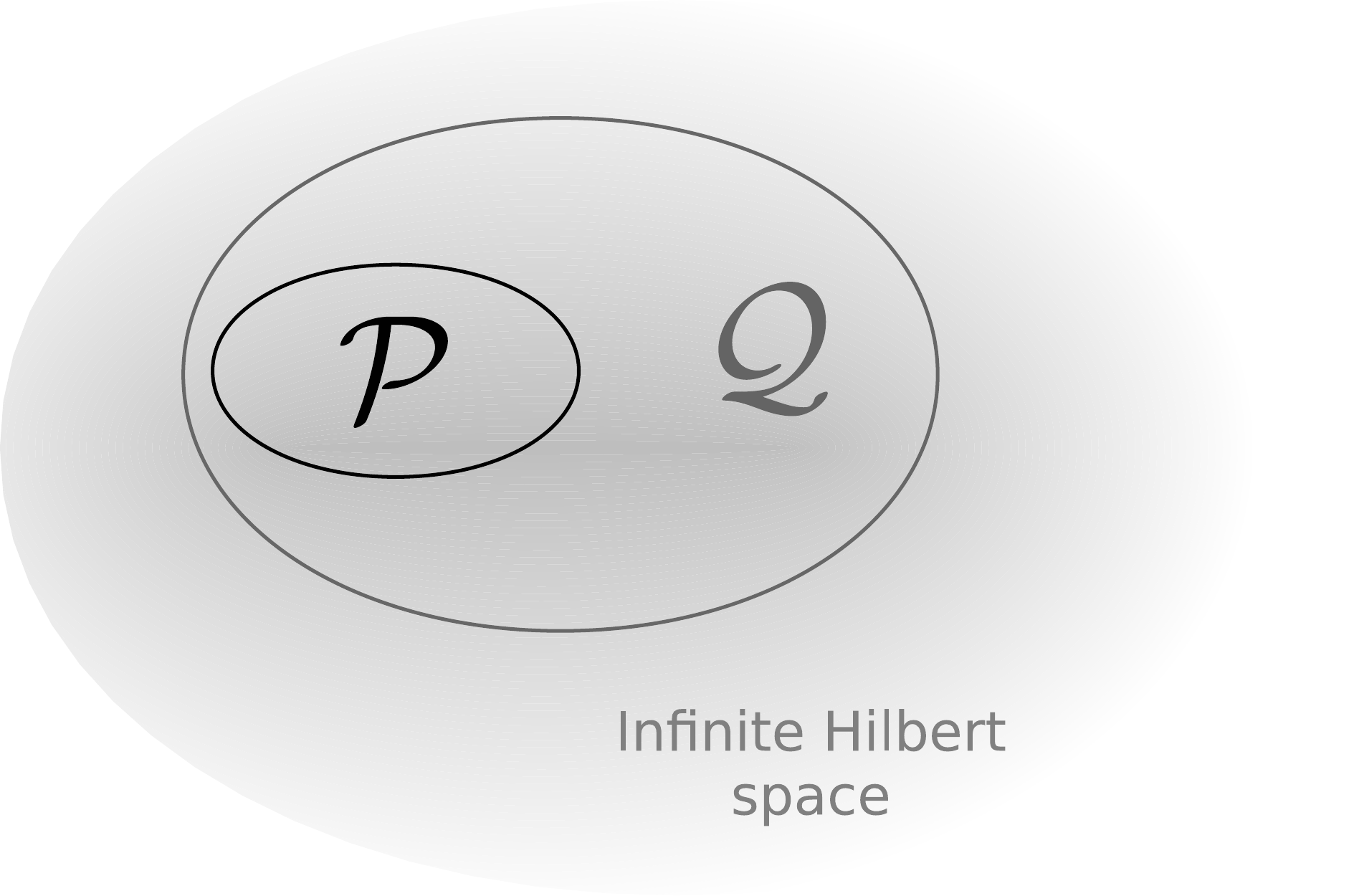}
\end{center}
\caption{Separation of the many-body space $\mathcal{S}$ spanned by the finite single-particle basis, into $\mathcal{P}\oplus\mathcal{Q}$.}
\label{f:Pspace}
\end{figure}

In theory the single-particle basis is infinite so that the exact state $\ket{\Psi_{ex}}$ does not depend on the nature of the orbitals, and the only unknown parameters to be determined are the mixing coefficients $\{A_\alpha\}$. Practically however, one has to work with finite spaces. Since nuclei are known to be very collective systems, the single-particle basis has to be large enough in order to approximate the exact solution to a good accuracy. Because the number of configurations grows combinatorially with the number of particles and single-particle states, and is drastically increased by the presence of two types of nucleons, it is most often impossible to perform exact calculations in the full configuration space $\mathcal{S}$ spanned by the (finite) one-body basis. Consequently, one is forced to restrict expansion (\ref{e:wf}) to configurations belonging to a subspace $\mathcal{P}$ of $\mathcal{S}=\mathcal{P}\oplus\mathcal{Q}$. This truncation is schematically represented in Fig.~\ref{f:Pspace}. 

\noindent Thus, the approximate nuclear state $\ket{\Psi}$ that one considers reads in fact
\begin{equation}
\ket{\Psi} = \sum_{\alpha\in \mathcal{P}} A_\alpha \ket{\phi_\alpha} \; .
\label{e:wf_trunc}
\end{equation}
Naturally the subspace $\mathcal{P}$ should be chosen in order to contain the physically most relevant many-body states. Different truncation schemes can be adopted: a "Shell-Model-type" truncation dividing the single-particle space into a frozen filled core, an active valence space and remaining empty orbits; a selection of the Slater determinants according to their excitation order $M_\alpha$ ($1p$-$1h$, $2p$-$2h$...), or according to their excitation energy $E_\alpha^*=E_\alpha - E_0$ from the uncorrelated ground-state $\ket{\phi_0}$. Combinations of these criteria can also be applied, but special care always needs to be taken in the choice of the configurations in order to ensure as much as possible the preservation of the fundamental symmetries of the nuclear Hamiltonian, as for example the rotational or the time-reversal invariances. In any case, such a drastic truncation now renders the wave function significantly dependent on the nature of the single-particle basis. The strategy of the MPMH configuration mixing approach is thus to determine the optimal set of single-particle states to be used to construct the selected many-body configurations. By enriching the restricted subspace $\mathcal{P}$, this procedure is expected to minimize the role of its orthogonal space $\mathcal{Q}$ on the calculation of observables. 
\\ 
\\ 
Consequently, the two sets of unknown parameters to be determined are: the mixing coefficients $\{A_{\alpha}\}$, as well as the single-particle orbitals $\{ \varphi_{i_{\tau}}, \tau=(\pi,\nu)\}$ used to build the many-body states. These quantities are obtained by applying a variational principle to the energy functional $\mathcal{E}[\Psi]=\braket{\Psi|\hat H|\Psi}$ of the system. The equation determining the weights $\{A_{\alpha}\}$ of the configurations is obtained by requiring $\mathcal{E}[\Psi]$ to be stationary with respect to infinitesimal variations $\delta A_{\alpha}^{*}$ of the coefficients, while the orbitals are kept fixed. Similarly the orbitals are optimized by fixing the coefficients and minimizing $\mathcal{E}[\Psi]$ with respect to the single-particle states $\{ \varphi_{i_{\tau}}\}$. This leads to the following system of coupled equations
\begin{numcases}{}
\delta_A \mathcal{E} [\Psi] =0 \label{e:syst1} \\
\delta_{\varphi} \mathcal{E} [\Psi] = 0 \; , \label{e:syst2}
\end{numcases}
where $\delta_A$ and $\delta_{\varphi}$ denote the variations with respect to the mixing coefficients and the orbitals, respectively.

\subsection{First variational equation: the mixing coefficients}
\begin{figure}[t]
\begin{center}\includegraphics[width=0.9\columnwidth]{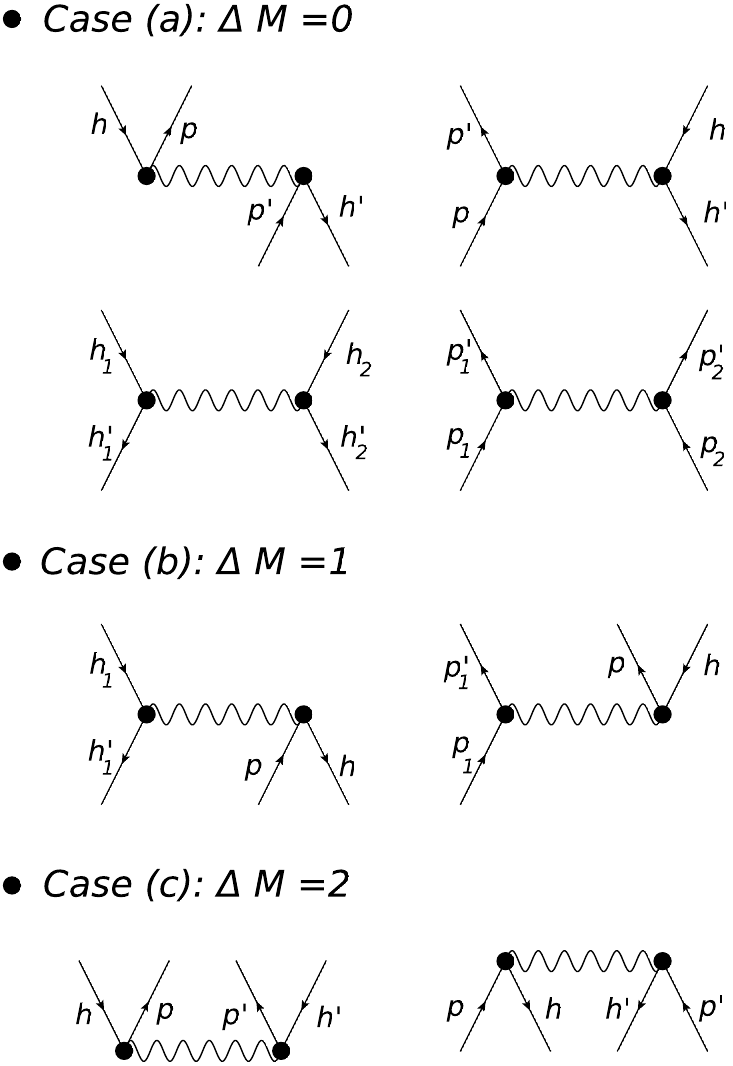}
\end{center}
\caption{Diagrams representing the matrix elements $\braket{\phi_\alpha|:\hat V^{2N}:|\phi_\beta}$, according to the difference in excitation orders $\Delta M=|M_\alpha - M_\beta|$.}
\label{f:vertex}
\end{figure}

Differentiating the energy with respect to the mixing coefficients, one finds the first extremum condition (\ref{e:syst1}) to be equivalent to the eigenvalue equation
\begin{equation}
\sum_{\beta \in \mathcal{P}} \braket{\phi_\alpha|\hat H|\phi_\beta} A_\beta = \lambda A_\alpha \; .
\label{e:eq1}
\end{equation}
Eq.~(\ref{e:eq1}) is common to Configuration Interaction (CI) methods like the SM, and represents the diagonalization of the Hamiltonian matrix 
in the many-body space $\mathcal{P}$. The nuclear states $\ket{\Psi}$ correspond to the eigenvectors of $\hat H$, while the eigenvalues $\lambda$ give the total energy of the system.
\\ 
\\
Eq.~(\ref{e:eq1}) introduces explicit correlations in the nuclear state $\ket{\Psi}$. As already largely discussed in \cite{Pillet1}, these correlations are of different physical types. Indeed, if for instance one neglects the three-body part of the residual interaction, the matrix elements $\braket{\phi_\alpha|:\hat V^{2N}:|\phi_\beta}$ can be represented by different types of vertices, as shown in Fig.~\ref{f:vertex}. According to the difference in excitation orders $\Delta M=|M_\alpha - M_\beta|$ of the two Slater determinants, these vertices can correspond to pairing correlations, RPA-type correlations which generate collective vibrations of the system, or particle-vibration-type correlations which couple the collective states to the single-particle motion. Although they are treated on the same footing, these correlations are restricted to the subspace $\mathcal{P}$ only, so that at this stage the subspace $\mathcal{Q}$ has been completely left ignored.
\\
\\
In order to partly make up for this truncation, the idea is now to find the set of single-particle orbitals which render $\mathcal{P}$ as physically relevant as possible.

\subsection{Second variational equation: the single-particle orbitals}
We now minimize the energy functional $\mathcal{E}[\Psi]$ with respect to the single-particle states. A variation of the creation operators $\{a^{\dagger}_i \}$ can be obtained from a general unitary transformation
\begin{equation}
a^{\dagger}_i \rightarrow e^{i\hat{T}} a^{\dagger}_i e^{-i\hat{T}}  =  a^{\dagger}_i + i\left[ \hat{T},a^{\dagger}_i \right] - 
\frac{1}{2} \left[\hat{T}, \left[\hat{T},a^{\dagger}_i \right] \right] +... \; ,
\label{e:transfo_unit}
\end{equation} 
where $\hat T$ is an infinitesimal hermitian one-body operator and $\left[, \right]$ denotes the commutator. The resulting first order variation of the wave function reads 
\begin{equation}
\ket{\delta \Psi}=i\hat T \ket{\Psi}, 
\end{equation}
so that the extremum condition (\ref{e:syst2}) finally becomes
\begin{equation}
\braket{\Psi |\left[ \hat H, \hat T \right]| \Psi} = 0 \; .
\label{e:Brillouin}
\end{equation}
This condition is often referred in the literature as "Generalized Brillouin equation" \cite{Brillouin1,Brillouin2}. Eq.~(\ref{e:Brillouin}) can also conveniently be recasted into the following general inhomogeneous mean-field equation
\begin{equation}
\left[ \hat{h}[\rho,\sigma] , \hat\rho \right] = \hat{G}[\rho,\sigma,\chi] \; .
\label{e:eq2}
\end{equation}
In Eq.~(\ref{e:eq2}) $\rho$ is the one-body density matrix of the correlated state:
\begin{equation}
\rho_{ij}=\braket{\Psi|a^{\dagger}_j a_i|\Psi} \; .
\end{equation}
The eigenbasis of $\rho$ is called "natural" basis, and its eigenvalues $\{n_i\}$ are occupation numbers. 
\\
The quantities $\sigma$ and $\chi$ denote the two- and three-body correlation matrices, respectively. They are defined by
\begin{equation}
\braket{\Psi|a^{\dagger}_1 a^{\dagger}_2 a_{2'} a_{1'}|\Psi} = (1-P_{12}) \rho_{1'1} \rho_{2'2} + \sigma_{11',22'} \; ,
\end{equation} 
and 
\begin{eqnarray}
&&\braket{\Psi| a^{\dagger}_1 a^{\dagger}_2 a^{\dagger}_3 a_{3'}a_{2'} a_{1'} |\Psi} = \nonumber \\
                   && \hspace{1cm} (1-P_{12}-P_{13})(1-P_{23}) \rho_{1'1} \rho_{2'2} \rho_{3'3} \nonumber \\
                   && \hspace{1cm} + (1-P_{12}-P_{13})\rho_{1'1}\sigma_{22',33'} \nonumber \\
                   && \hspace{1cm} + (1-P_{12}-P_{23})\rho_{2'2}\sigma_{11',33'} \nonumber \\
                   && \hspace{1cm} + (1-P_{13}-P_{23})\rho_{3'3}\sigma_{11',22'} \nonumber \\
                   && \hspace{1cm} + \chi_{11',22',33'} \; ,
\end{eqnarray} 
where the set of $P_{ij}$ represents two-nucleon exchange operators.
\\
The one-body mean-field Hamiltonian $h[\rho,\sigma]$ is defined as
\begin{eqnarray} 
 h[\rho,\sigma]_{ij} &\equiv& K_{ij}+\Gamma^{2N}[\rho]_{ij}+\Gamma^{3N}[\rho,\sigma]_{ij} \nonumber \\
                           &=& K_{ij} + \sum_{kl} \widetilde{V}^{2N}_{ikjl} \rho_{lk} \nonumber \\
                           &&  + \frac{1}{4} \sum_{klmn} \widetilde{V}^{3N}_{ikl,jmn} \braket{\Psi|a^{\dagger}_k a^{\dagger}_l a_n a_m|\Psi}  \nonumber \\
                           &=& K_{ij} + \sum_{kl} \widetilde{V}^{2N}_{ikjl} \rho_{lk} \nonumber \\
                           && + \frac{1}{2} \sum_{klmn} \widetilde{V}^{3N}_{ikl,jmn} \rho_{mk}\rho_{nl} \nonumber \\
                           && + \frac{1}{4} \sum_{klmn} \widetilde{V}^{3N}_{ikl,jmn} \sigma_{km,ln} \;.
\label{e:mean_field}                           
\end{eqnarray}
The eigenstates of $h[\rho,\sigma]$ constitute the "canonical" basis, and its eigenvalues $\{\varepsilon_\mu\}$ are single-particle energies.
\\
Finally the source term $G[\rho,\sigma,\chi]$ contains the effect of two- and three-body correlations beyond the mean-field $h[\rho,\sigma]$. It is an anti-hermitian quantity which can be written as
\begin{equation}
G[\rho,\sigma,\chi] = F[\rho,\sigma,\chi] - F^\dagger[\rho,\sigma,\chi] \; ,
\label{e:G_3N}
\end{equation}
with 
\begin{eqnarray}
F[\rho,\sigma,\chi]_{ij} &=& F^{2N}[\sigma]_{ij} +F^{3N}[\rho,\sigma,\chi]_{ij} \nonumber \\
                         &=& \frac{1}{2}  \sum_{klm} \sigma_{ki,lm} \widetilde{V}^{2N}_{kl,jm}    \nonumber \\
                         && + \frac{1}{2} \sum_{klmnq} \sigma_{ki,mn} \widetilde{V}^{3N}_{klm,nqj} \rho_{ql}  \nonumber \\
                         && + \frac{1}{12} \sum_{klmnq} \chi_{kn,lq,mi} \widetilde{V}^{3N}_{klm,nqj} \; .
\label{e:F_3N}            
\end{eqnarray}

\begin{figure}[t]
\begin{center}
\includegraphics[width=0.6\columnwidth]{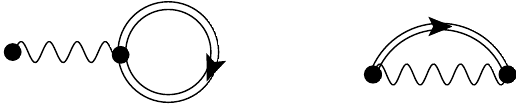}
\caption{Diagrammatic representation of the direct (left) and exchange (right) parts of the average potential $\Gamma^{2N}[\rho]$. The double line denotes the correlated one-body density.}
\label{f:Gamma}
\end{center}
\end{figure}

\begin{figure}[t]
\begin{center}
\includegraphics[width=0.95\columnwidth]{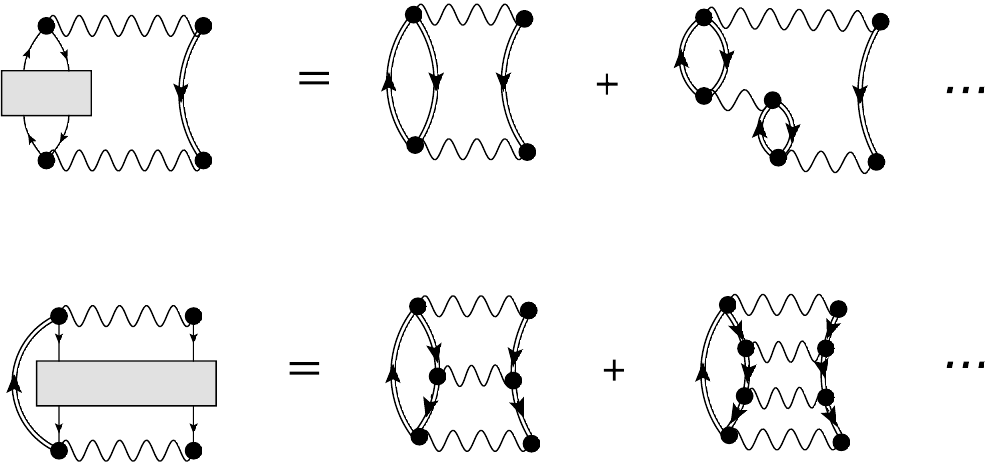}
\caption{Resummation of ring (top) and ladder (bottom) diagrams in $G^{2N}[\sigma]$.}
\label{f:Gsig}
\end{center}
\end{figure}

In fact one can show that the orbital equation (\ref{e:eq2}) can alternatively be derived from the formalism of Green's functions at equal times and more precisely, from the equation of motion relating the one-body propagator to the two-body propagator (see appendix \ref{a1}). An identification with the Dyson equation shows that the average potential $\Gamma[\rho,\sigma]\equiv \Gamma^{2N}[\rho]+ \Gamma^{3N}[\rho,\sigma]$ in Eq.~(\ref{e:mean_field}) corresponds to the time-independent part of the full one-nucleon self-energy. It is represented in Fig.~\ref{f:Gamma} in the case of a two-body interaction only. Additionally, the source term $G[\rho,\sigma,\chi]$ can be related to the dynamical part of the self-energy through a certain equal time limit (see appendix \ref{a1}). This source term contains the resummation of many diagrams related to the various types of correlations contained in the correlated wave function $\vert \Psi \rangle$, as illustrated in Fig.~\ref{f:Gsig}. Moreover, since the configuration mixing is performed in such a way to preserve important symmetries of the nuclear Hamiltonian, as the ones associated to the particle number, the total angular momentum or the time-reversal invariance, the source term $G[\rho,\sigma,\chi]$ also contains the correlations related to these symmetry preservations.

\noindent This analysis confirms that, whenever the densities are calculated with the \textit{exact} wave function of the system, Eq.~(\ref{e:eq2}) is automatically fulfilled. The mean-field defined in Eq.~(\ref{e:mean_field}) is the most general mean-field that can be constructed considering a three-body Hamiltonian (\ref{e:H_3N}). The part $\Gamma^{2N}[\rho]$ implicitly couples back to the correlations of the system by averaging the two-body interaction over both hole and particle states -- contrary to a Hartree-Fock field which averages over orbits under the Fermi level only. Moreover the part $\Gamma^{3N}[\rho,\sigma]$ introduces an explicit dependence on the two-body correlation matrix by averaging the three-body force with the full two-body density. It has been extensively discussed in \cite{Baranger,DuguetHagen} that the eigenvalues of such a general mean-field constitute the most unambiguous definition of single-particle energies, and physically correspond to centroids of one-nucleon separation energies. The theory of the general mean-field has also been widely exposed in \cite{BookBlaizot} from the point of view of perturbation theory. The authors emphasize the importance of consistency between mean field and correlations in order for the theory to be variational.

\subsubsection*{Role and interpretation of the orbital equation}
\begin{figure}[t]
\begin{center}\includegraphics[width=0.8\columnwidth]{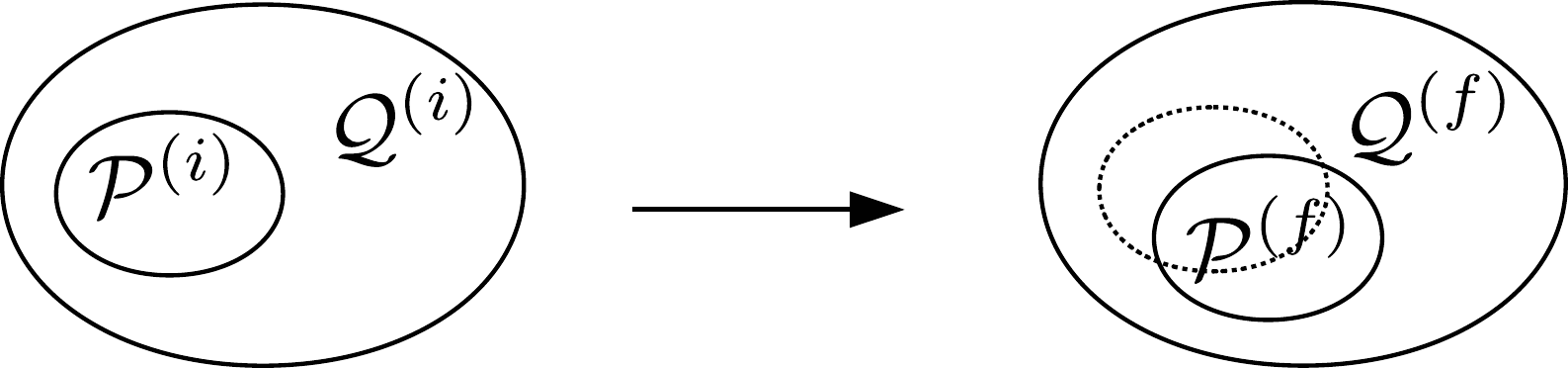}
\end{center}
\caption{Modification of the subspaces $\mathcal{P}$ and $\mathcal{Q}$ via the optimization of orbitals.}
\label{f:PQmodif}
\end{figure}

In practice, when truncations are applied to the wave function $\vert \Psi \rangle$, the orbital equation (\ref{e:eq2}) not only allows to achieve consistency between mean-field and correlation content, it is also expected to help compensate (partly) for these truncations. Let us justify this point. 
\\
The nuclear state $\ket{\Psi}$ resulting from the diagonalization (\ref{e:eq1}) of the many-body Hamiltonian is restricted to a selected subspace $\mathcal{P}$ of the full Hilbert space, i.e. $\ket{\Psi}=\hat P \ket{\Psi}$, where $\hat P$ is the projector onto $\mathcal{P}$. The variation $\ket{\delta \Psi}$ that was obtained from the transformation (\ref{e:transfo_unit}) of single-particle states can be divided into a part in $\mathcal{P}$ and a part which belongs to the orthogonal subspace $\mathcal{Q}$ as
\begin{equation}
\ket{\delta \Psi} = \hat P \ket{\delta \Psi} + \hat Q \ket{\delta \Psi} \equiv \ket{\delta \Psi}_{\mathcal{P}} + \ket{\delta \Psi}_{\mathcal{Q}} \; .
\end{equation} 
Thus, the corresponding variation of the energy can also be written as
\begin{eqnarray}
\delta_{\varphi} \mathcal{E}[\Psi] &=&{\bra{\Psi}} \hat{H} \ket{\delta \Psi} +  \braket{\delta \Psi| \hat{H} |\Psi}\nonumber \\
                                   &=& {\bra{\Psi}} \hat{H} \ket{\delta \Psi}_{\mathcal{P}} 
                                               +\prescript{}{\mathcal{P}}{\bra{\delta \Psi}} \hat{H} \ket{ \Psi} \nonumber \\
                                   && \hspace{0.5cm}  +{\bra{\Psi}} \hat{H} \ket{\delta \Psi}_{\mathcal{Q}} 
                                               +\prescript{}{\mathcal{Q}}{\bra{\delta \Psi}} \hat{H} \ket{ \Psi}\nonumber \\
                                   &=& {\bra{\Psi}} \hat{P}\hat{H}\hat{P} \ket{\delta \Psi}
                                               +{\bra{\delta \Psi}}\hat{P} \hat{H}\hat{P} \ket{ \Psi} \nonumber \\
                                   && \hspace{0.5cm} +{\bra{\Psi}} \hat{P}\hat{H}\hat{Q} \ket{\delta \Psi}
                                               +{\bra{\delta \Psi}} \hat{Q}\hat{H}\hat{P} \ket{ \Psi} \; . \nonumber \\
\label{e:PQcoupling}                                               
\end{eqnarray}
One notices that this first order variation does not allow to account for terms in $\hat{H}_{\mathcal{Q}\mathcal{Q}} \equiv \hat{Q}\hat{H}\hat{Q}$ representing propagation in the subspace $\mathcal{Q}$. However, couplings between $\mathcal{P}$ and $\mathcal{Q}$ appear through $\hat{H}_{\mathcal{P}\mathcal{Q}} \equiv \hat{P}\hat{H}\hat{Q}$ and $\hat{H}_{\mathcal{Q}\mathcal{P}} \equiv \hat{Q}\hat{H}\hat{P}$.
\\ 
\\
Another argument can be made by considering the optimal single-particle states solution of Eq.~(\ref{e:eq2}). Starting from an initial arbitrary set of single-particle states $\{a^{\dagger}\}$, one can define an initial subspace $\mathcal{P}$, denoted as $\mathcal{P}^{(i)}$, according to a certain truncation scheme. The orbital equation then leads to a new set $\{b^{\dagger}\}$ that can be expressed as
\begin{eqnarray}
 b^{\dagger}_i = e^{i\hat\Lambda} a^{\dagger}_i e^{-i\hat\Lambda} =\sum_{j}  a^{\dagger}_j  \left( e^{i\hat\Lambda} \right)_{ji} \equiv \sum_{j}  
 a^{\dagger}_j  \theta_{ji} \; ,
\end{eqnarray} 
where the sum runs over all states $j$ (of same symmetry than $i$ in a symmetry-conserving approach), and $\hat \Lambda=\sum_{kl} \Lambda_{kl} a^\dagger_k a_l $ is the one-body operator parametrizing the unitary transformation. Under this transformation, the many-body configurations therefore vary as
\begin{eqnarray}
|\phi_{\alpha}\rangle \rightarrow |\phi'_{\alpha}\rangle 
                          &=& e^{i\hat\Lambda} |\phi_{\alpha}\rangle \nonumber \\
                          &=&  |\phi_{\alpha}\rangle + i \sum_{ij} \Lambda_{ij} a^{\dagger}_i a_j  |\phi_{\alpha}\rangle \nonumber \\
                          &&   -\frac{1}{2} \sum_{ijkl} \Lambda_{ij} \Lambda_{kl} a^{\dagger}_i a_j  a^{\dagger}_k a_l |\phi_{\alpha}\rangle +... \; .
\label{e:mpmh_eq2}                                                                                            
\end{eqnarray} 
Mixing the single-particle states thus amounts to creating multiparticle-multihole excitations on top of the existing configurations. These multiparticle-multihole excitations extend to the whole (finite) single-particle basis one is considering, and are not restricted to a certain valence space or to a maximum excitation order -- Of course the contribution of configurations with high excitation order or high excitation energy are expected to be small if the initial single-particle states and selection scheme are physically relevant. The role of the optimization of orbitals is to produce an optimal final space $\mathcal{P}^{(f)}$, combination of the initial $\mathcal{P}^{(i)}$ and $\mathcal{Q}^{(i)}$, so that the influence of $\mathcal{Q}^{(f)}$ on the description of the nuclear state is minimized (see Fig.~\ref{f:PQmodif}).  In section \ref{Results}, we will see to what extent the two initial spaces are mixed, according to the correlation content of the system under study. Let us emphasize that, since it acts at the one-body level, the transformation of single-particle states does not create additional correlations. $\Lambda$ being a one-body operator, the excitations in Eq.~(\ref{e:mpmh_eq2}) are always built as products of $1p$-$1h$ excitations. Thus one should not expect that the orbital equation will fully make up for the truncation of the wave function. \\ \\

We end this section on a final remark about the interpretation of the orbital equation. Eq.~(\ref{e:eq2}) is a generalization of the Hartree-Fock equation when correlations are introduced. In the limit where only the reference state is included in expansion (\ref{e:wf}), i.e. when $\ket{\Psi}=\ket{\phi_0}$, one gets back the usual Hartree-Fock equation $\left[ h[\rho_0],\rho_0 \right] = 0$ where $\rho_0=\braket{\phi_0|\hat \rho|\phi_0}$ is a step function. Such a commutation property ensures the existence of a basis diagonalizing both $h[\rho_0]$ and $\rho_0$. This eigenbasis defines the Hartree-Fock single-particle states one seeks. In the general case where a configuration mixing is introduced, $\left[ h[\rho],\rho \right] = G[\sigma] \ne 0$. The "canonical" and "natural" bases do not coincide anymore and one cannot define states with definite single-particle energies and occupation numbers at the same time. The question of which one is the optimal basis then arises. In this approach we look for the single-particle states that are used to construct the configurations included in the (truncated) expansion of $\ket{\Psi}$. While the mean-field is rather related to the energy of the system, the density contains direct information about the content of the wave function. Therefore it seems legitimate to choose the natural orbitals as optimal orbitals. In this way, the reference state $\ket{\phi_0}$, which is then obtained by filling the orbitals with the higher occupations $n_i$, will incorporate a greater content of the wave function and will approximate at best the correlated state $\ket{\Psi}$. Hence, the natural basis satisfying Eq.~(\ref{e:eq2}) is expected to minimize the correlations in the sense that the weight of the reference state $|A_0|^2$ in the configuration mixing should be maximized.


\section{Applications with the D1S Gogny force}
\label{section2}
Although we derived the formalism in a general context, the applications in this work are performed without an explicit three-body force. 
We use instead the two-body D1S Gogny interaction \cite{D1S}. This phenomenological force contains a density-dependent term which effectively 
accounts for many-body effects as well as short-range correlations. Let us note that this interaction was originally created in order to perform 
mean-field calculations or reasonable extensions, such as RPA (Random Phase Approximation) or GCM (Generator Coordinate Method), for which it has shown globally successful results \cite{BlaizotGogny,GognyPadjen,BlaizotBerger,Peru1,Donno1,Donno2,Peru2,Peru3,Delaroche,Robledo,Rodriguez}. 
Nothing guarantees however that it will be adapted to the simultaneous treatment of all types of long-range correlations, in the various spin(S)-isospin(T) channels. The results obtained will depend on that. Moreover, the question of which density is to be used in the interaction remains open when going to correlated systems. In the following we use the density built with the correlated state $\ket{\Psi}$. This is a prescription which has the advantage of simplifying the variational equations. However, since the phenomenological nature of the Gogny force makes impossible to disentangle what effects are already included in the interaction, uncontrolled over-counting effects are likely to occur. Finally, we are fully aware that, in an ultimate step, this kind of approach requires a fully finite range Gogny interaction. Work is done in this direction \cite{D2,PilletPena}.

\subsection{Variational equations}
Since the D1S Gogny force is density-dependent, the variational principle does not reduce to the equations of the previous section obtained by putting the terms in $V^{3N}$ to zero. In fact new derivatives appear \cite{Pillet1} and it is easily shown that the two variational equations to be solved become
\begin{numcases}{}
\sum_{\beta} A_{\beta} \braket{\phi_{\alpha}|\hat{\mathpzc{H}}[\rho,\sigma] |\phi_{\beta}} = \lambda A_{\alpha}, \;\; \forall \alpha  \label{e:eq1_Gogny} \\
\left[ \hat{\mathpzc{h}}[\rho,\sigma] , \hat{\rho} \right] = \hat{G}[\sigma] \; . \label{e:eq2_Gogny}
\end{numcases}
In Eq.~(\ref{e:eq1_Gogny}), the Hamiltonian matrix to be diagonalized has been modified and now reads
\begin{eqnarray}
\hat{\mathpzc{H}}[\rho,\sigma] &=& \hat{H}^{2N}[\rho] + \hat{\mathpzc{R}}[\rho,\sigma] \nonumber \\
                               &=& \hat K + \hat{V}^{2N}[\rho] + \hat{\mathpzc{R}}[\rho,\sigma] \; ,
\label{e:GognyHamilt}                                                              
\end{eqnarray}
where
\begin{eqnarray}
\hat{\mathpzc{R}}[\rho,\sigma] &=& \int \D^3 r \braket{\Psi| \frac{\delta \hat{V}^{2N}[\rho]}{\delta \rho(\vec{r})} |\Psi} \hat{\rho}(\vec{r})
                                    \nonumber \\
                               &=& \frac{1}{4} \int \D^3 r \sum_{klmn} \braket{kl| \frac{\delta \widetilde{V}^{2N}[\rho]}{\delta \rho(\vec{r})}|mn} 
                                     \nonumber \\
                               &&  \hspace{0.7cm} \times \braket{\Psi|a^{\dagger}_k a^{\dagger}_l a_n a_m|\Psi} \hat{\rho}(\vec{r}) 
                                      \nonumber \\
                               &=& \frac{1}{4} \int \D^3 r \sum_{klmn} \braket{kl| \frac{\delta \widetilde{V}^{2N}[\rho]}{\delta \rho(\vec{r})}|mn}
                                    \nonumber \\ 
                               && \hspace{0.7cm} \times \left( \rho_{mk} \rho_{nl} - \rho_{ml} \rho_{nk} + \sigma_{km,ln} \right) \hat{\rho}(\vec{r}) \; .
                               \nonumber \\
\end{eqnarray}
The operator $\hat{\mathpzc{R}}[\rho,\sigma]$ is called "rearrangement term" and represents the response of the system to a variation of density. 
Although one-body operator, $\hat{\mathpzc{R}}[\rho,\sigma]$ depends on the two-body correlation matrix $\sigma$ and thus requires the computation of this complicated quantity. The dependence of $\mathpzc{H}$ on the one- and two-body densities of the system renders Eq.~(\ref{e:eq1_Gogny}) non-linear. A solution of this equation cannot be obtained via a single diagonalization anymore but requires an iterative procedure. Moreover the eigenvalues $\lambda$ of $\mathpzc{H}[\rho,\sigma]$ no longer correspond to the energies of the system under study. In fact we have $E[\Psi]=\braket{\Psi|\hat H^{2N}[\rho]|\Psi}= \lambda - \braket{\Psi|\hat{\mathpzc{R}}[\rho,\sigma]|\Psi}$. We also note that, in both Eqs.~(\ref{e:eq1_Gogny}) and (\ref{e:eq2_Gogny}), all the direct and exchange terms due to the Pauli principle, are treated exactly.
\\
\\ 
In the orbital equation (\ref{e:eq2_Gogny}), the mean-field Hamiltonian is modified as
\begin{eqnarray}
\mathpzc{h}_{ij}[\rho,\sigma]  &=& h_{ij}[\rho] + \mathpzc{R}_{ij}[\rho,\sigma] \nonumber \\
                               &=& K_{ij} + \Gamma_{ij}[\rho] + \mathpzc{R}_{ij}[\rho,\sigma] \nonumber \\
                               &=& K_{ij} + \sum_{kl} \braket{ik|\widetilde{V}^{2N}[\rho]|jl} \rho_{lk}+ \mathpzc{R}_{ij}[\rho,\sigma] \; . \nonumber \\
\end{eqnarray}
Expressing the rearrangement term as
\begin{equation}
\mathpzc{R}_{ij}[\rho,\sigma] = \frac{1}{4} \sum_{klmn} \braket{kl|\frac{\partial \widetilde{V}^{2N}[\rho]}{\partial \rho_{ji}}|mn}  
                                \braket{\Psi| a^{\dagger}_k a^{\dagger}_l a_n a_m| \Psi} \; ,
\end{equation}
one notes the similarity between $\mathpzc{R}[\rho,\sigma]$ and the potential $\Gamma^{3N}[\rho,\sigma]$ appearing in the mean-field (\ref{e:mean_field}) that was derived from an explicit three-body Hamiltonian. Thus one can say that the $\rho$-dependency of the two-body force allows to simulate the part of the three-body interaction that is averaged on two particles. Of course, the D1S force being phenomenological, higher many-body effects are also implicitly accounted for and no formal link can be made.
\\
\\
The source term of the orbital equation now reads
\begin{eqnarray}
G[\sigma]_{ij} &=&  \frac{1}{2} \sum_{klm} \sigma_{ki,lm} \braket{kl|\widetilde{V}^{2N}[\rho]|jm} \nonumber \\
               &&  - \frac{1}{2}  \sum_{klm} \braket{ik|\widetilde{V}^{2N}[\rho]|lm}  \sigma_{jl,km}  \; .
\label{e:G_Gogny}                        
\end{eqnarray}
\\
Comparing (\ref{e:G_Gogny}) to the source term (\ref{e:G_3N}) derived in the previous section, we note that only the part derived from the two-body interaction is reproduced. While it is important for the mean-field description, neglecting the three-body part of the residual interaction is usually a reasonable approximation.
\\ \\
Finally, because the Gogny force has been fitted to experimental data, it empirically accounts for part of the subspace $\mathcal{Q}$ that was discussed in section \ref{section1}. Therefore the division of the many-body space in terms of  $\mathcal{P}$ and $\mathcal{Q}$ is not clear when using such a density-dependent interaction, and the rearrangement terms are likely to reduce the effect of the orbital equation.

\subsection{Numerical algorithm}
\begin{figure}[t]
\begin{center}
\includegraphics[width=0.6\columnwidth]{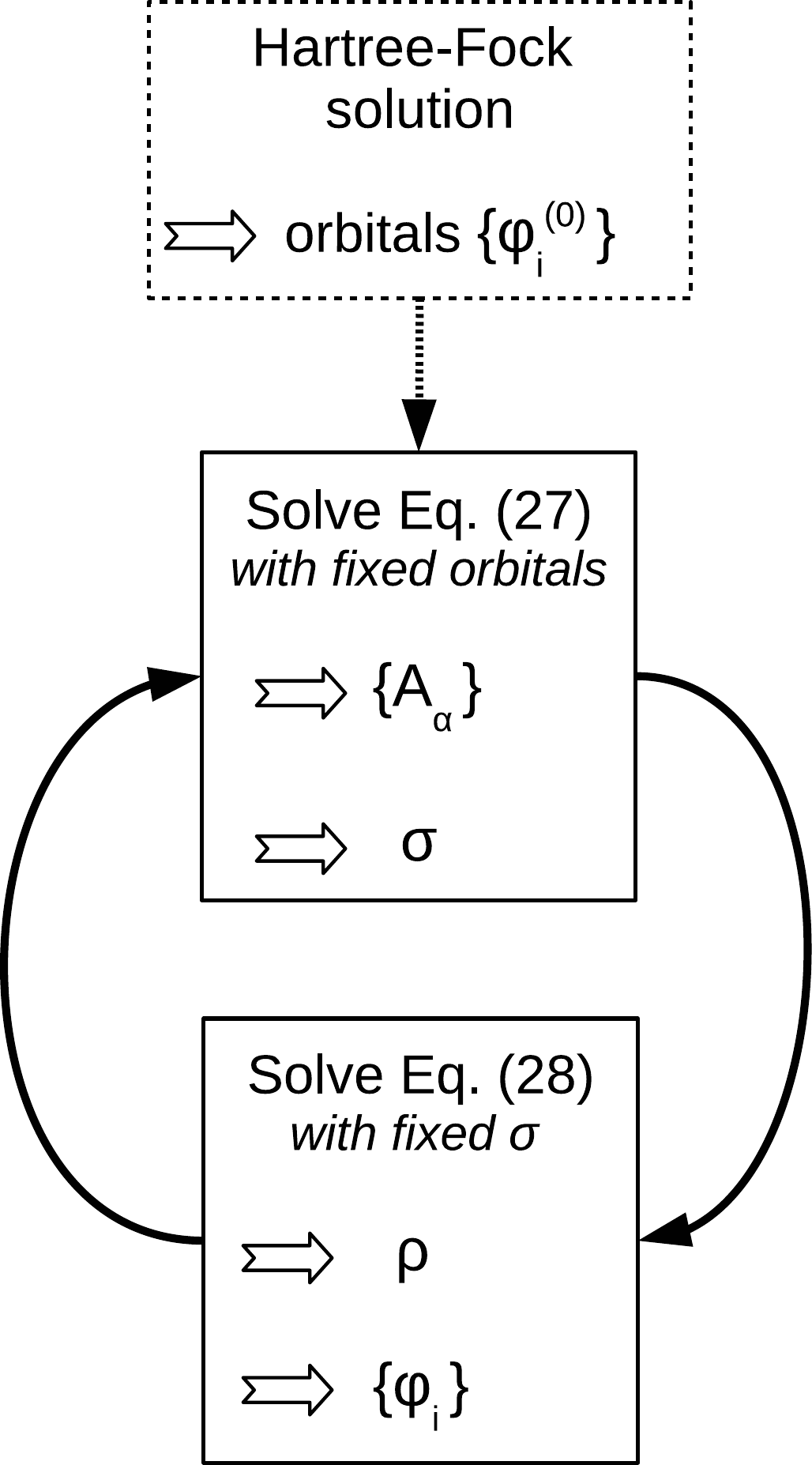}
\caption{Convergence procedure of the MPMH configuration mixing method.}
\label{f:global_algo2}
\end{center}
\end{figure}

Since the choice of orbitals $\{\varphi_i\}$ influences the mixing coefficients $\{A_\alpha\}$, and {\it vice versa}, Eqs.~(\ref{e:eq1_Gogny}) and (\ref{e:eq2_Gogny}) are coupled. A fully self-consistent solution can therefore be obtained via an iterative procedure where both equations are solved successively at each step of the process. More precisely, the global scheme that is adopted in this work is the following:
\begin{enumerate}
 \item Start by assuming a single-configuration wave function $\ket{\Psi^{(0)}}=\ket{\phi_0}$ so that no correlations are present, i.e. $\sigma^{(0)}=0$, and $\rho^{(0)}=\rho_0=\braket{\phi_0|\hat{\rho}|\phi_0}=\rho_0^2$. Solve the corresponding Eq.~(\ref{e:eq2_Gogny}): $\left[\mathpzc{h}[\rho^{(0)}],\rho^{(0)}\right]=0$. This is a standard Hartree-Fock calculation which leads to a first set of orbitals $\{\varphi_i^{(0)}\}$.
 \item \label{step2} Construct the many-body configurations $\{\phi_\alpha^{(0)}\}$ on these initial orbitals and solve Eq.~(\ref{e:eq1_Gogny}) to obtain a first set of ground state components $\{A_\alpha^{(1)}\}$. The correlation matrix $\sigma^{(1)}$ can then be calculated from these.
 \item Keeping $\sigma^{(1)}$ fixed, solve Eq.~(\ref{e:eq2_Gogny}), i.e. solve $\left[\mathpzc{h}[\rho^{(1)},\sigma^{(1)}],\rho^{(1)}\right] = G[\sigma^{(1)}]$, to obtain the one-body density $\rho^{(1)}$. The new single-particle states $\{\varphi^{(1)}_i\}$ are taken as eigenvectors of the solution $\rho^{(1)}$.
 \item Go back to step \ref{step2} using these new orbitals, and repeat the procedure until convergence is reached.
\end{enumerate}
This process is illustrated in Fig.~\ref{f:global_algo2}.
In principle convergence of both orbitals and mixing coefficients -- or equivalently of both one- and two-body densities -- must be checked. In the present work, the convergence criterion is however only set on the one-body density $\rho$. Convergence is assumed to be reached when variations of the density matrix $\Delta \rho_{ij} = |\rho_{ij}^{(N)} - \rho_{ij}^{(N-1)}|$ between two consecutive iterations $N-1$ and $N$, are smaller than $1\times10^{-4}$. In practice we observe that when convergence on $\rho$ is reached, $\sigma$ has converged to a similar accuracy.

\subsubsection{Solution of the first variational equation (\ref{e:eq1_Gogny})} 
Solving the eigenvalue equation (\ref{e:eq1_Gogny}) already represents a very difficult task in the context of SM methods because it involves the diagonalization of huge matrices. Moreover the implementation of proton-neutron contributions increases the computational challenge. Here, because of the $\rho$- and $\sigma$-dependence of the matrix $\mathpzc{H}[\rho,\sigma]$ to diagonalize, Eq.~(\ref{e:eq1_Gogny}) is in addition non-linear. A converged solution of this equation (with fixed orbitals) can in principle be achieved by iterating the diagonalization of $\mathpzc{H}[\rho,\sigma]$ until the mixing coefficients $\{A_\alpha\}$ have converged. However, when the size of the matrix is big, this numerical procedure can become very time consuming. Since Eq.~(\ref{e:eq2_Gogny}) will modify the orbitals and thus the mixing coefficients, it is not worth converging the coefficients $\{A_\alpha\}$ to a precise accuracy at this intermediate stage. Therefore, in concrete terms, we choose not to perform this sub-iterative process and to move to Eq.~(\ref{e:eq2_Gogny}) after one-single diagonalization. Ultimately, convergence will be reached through the global procedure. The diagonalization of the large Hamiltonian matrix is achieved using the numerical techniques developed by E. Caurier for large-scale SM calculations \cite{SM}.

\subsubsection{Solution of the second variational equation (\ref{e:eq2_Gogny})}
Solving the orbital equation (\ref{e:eq2_Gogny}) is far from being straightforward, and one could imagine different approaches. The idea followed in this work is to express the source term $G[\sigma]$ as a commutator with $\rho$, in order to obtain an homogeneous equation. Following this path one can show that Eq.~(\ref{e:eq2_Gogny}) can equivalently be expressed as
\begin{equation}
\left[ \hat{\mathpzc{h}}[\rho,\sigma] -\hat{Q}[\rho,\sigma] , \hat \rho \right] = 0 \; ,
\label{e:eq2_Gogny_hom}
\end{equation}
where we have defined what we call a "correlation field" $Q[\rho,\sigma]$. In the natural basis $\hat \rho \ket{i} = n_i \ket{i}$ this correlation field 
is given by
\begin{equation}
Q_{ij}[\rho,\sigma]=
\left\{
\begin{array}{ll}
 \frac{G_{ij}[\sigma]}{n_{j}-n_{i}} & \text{  if } n_{i}\ne n_{j}\\
 0                                  & \text{  otherwise. }
\end{array}
\right.
\end{equation} 
Eq.~(\ref{e:eq2_Gogny_hom}) now resembles some sort of Hartree-Fock equation where the mean-field $\mathpzc{h}[\rho,\sigma]$ is modified by the effect of two-body correlations through $Q[\rho,\sigma]$. The optimal single-particle basis that we seek is the one diagonalizing $\widetilde{\mathpzc{h}} \equiv \mathpzc{h}-Q$ and $\rho$ simultaneously. Since $\widetilde{\mathpzc{h}}[\rho,\sigma]$ depends on the solution $\rho$, this is of course a non-linear problem which can again be solved iteratively. Eq.~(\ref{e:eq2_Gogny_hom}) is solved with a fixed correlation content $\sigma$ (output of Eq.~(\ref{e:eq1_Gogny})), using the following algorithm:

\begin{enumerate}
 \item Start from an initial correlated one-body density $\rho=\rho_{init}$, and diagonalize it to obtain occupation numbers $\{n_i\}$.
 \item \label{stepb} Calculate and diagonalize $\widetilde{\mathpzc{h}}[\rho,\sigma]=\mathpzc{h}[\rho,\sigma] - Q[\rho,\sigma]$. The resulting eigenvectors  constitutes new single-particle states.
 \item Redistribute the particles on this new basis to obtain a new density $\rho$. 
 \item Go back to step \ref{stepb}... and so on until the $\rho$-matrix has converged in a given basis to the desired accuracy -- $1.0\times10^{-4}$ in this work. We note that in some cases however, there is no need to -- or it is even better not to -- completely converge this micro-process before going back to the first equation.
\end{enumerate}

\begin{figure}[t]
\begin{center}
\includegraphics[width=\columnwidth]{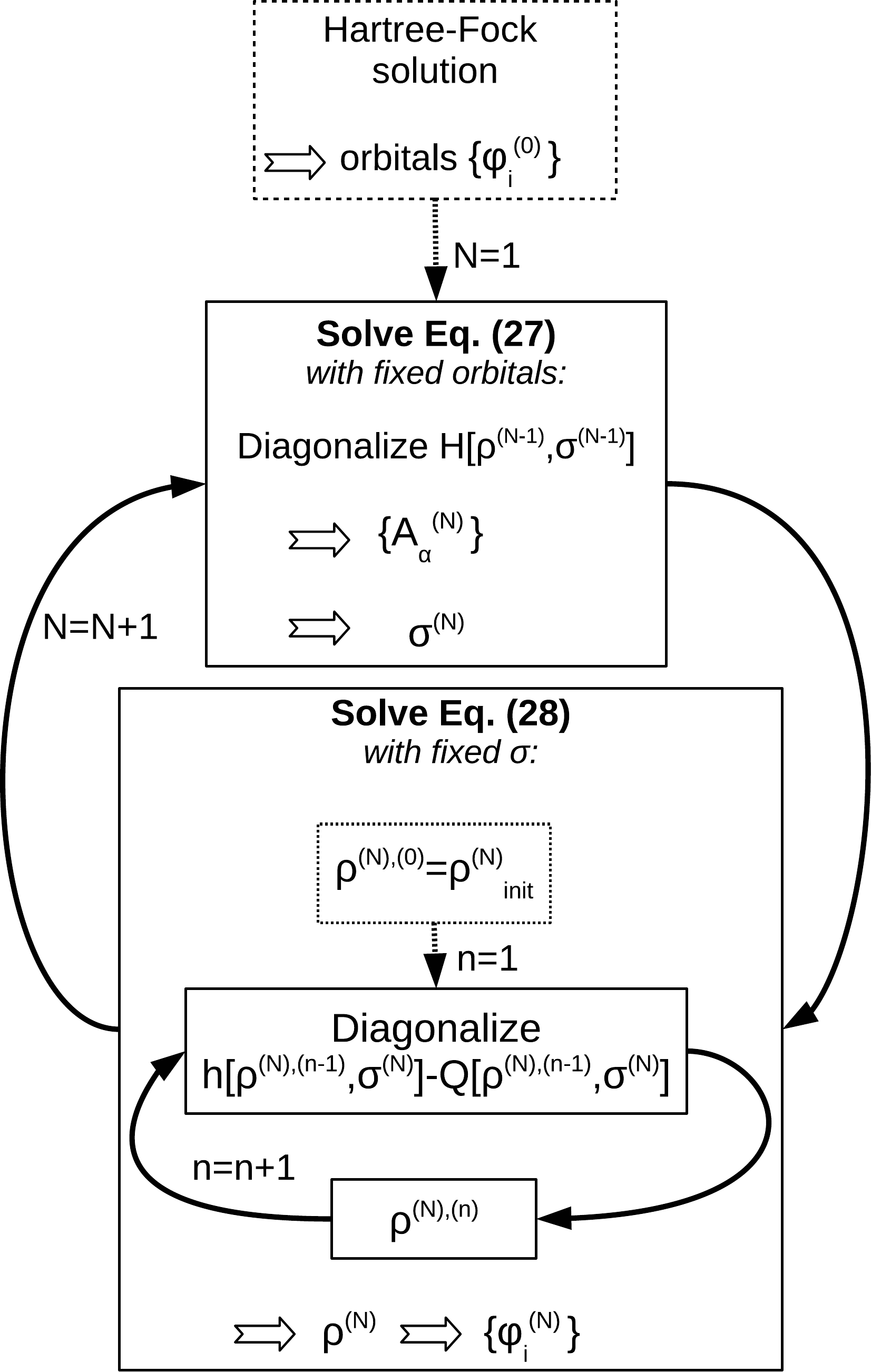}
\caption{Detailed convergence procedure of the MPMH configuration mixing method.}
\label{f:global_algo_detailed}
\end{center}
\end{figure}

We remind that this sub-convergence process takes place in a global iterative procedure. In order to differentiate both types of iterations, we call "macro-iteration" ($N$) an iteration of the global procedure and "micro-iteration" ($n$) an iteration of the process for solving  Eq.~(\ref{e:eq2_Gogny}) (or (\ref{e:eq2_Gogny_hom})) with fixed $\sigma$. The detailed global convergence procedure is shown in Fig.~\ref{f:global_algo_detailed}. 

\noindent The initial density $\rho_{init}^{(N)}$, starting point of the micro-convergence procedure, is taken as the density calculated from the solution of Eq.~(\ref{e:eq1_Gogny}), i.e. calculated as
\begin{equation}
\rho_{init,ij}^{(N)} = \sum_{\alpha\beta} A_{\alpha}^* A_{\beta} \braket{\phi_\alpha |a^\dagger_j a_i| \phi_\beta} \; .
\end{equation}
Ultimately, when full self-consistency is reached, the densities $\rho_{init}^{(N)}$ and $\rho^{(N)}$, output of Eqs.~(\ref{e:eq1_Gogny}) and (\ref{e:eq2_Gogny}) respectively, become identical. We therefore set the following convergence criteria for the global process:
$\left| \rho_{init}^{(N-1)}-\rho_{init}^{(N)} \right| \leqslant 1\times10^{-4}$, $\left| \rho^{(N-1)}-\rho^{(N)} \right| \leqslant 1\times10^{-4}$ and $\left| \rho^{(N)}-\rho_{init}^{(N)} \right| \leqslant 1\times10^{-4}$.


\section{Results for $^{12}$C} 
\label{Results}
\begin{figure}[t]
\begin{center}
\includegraphics[width=\columnwidth]{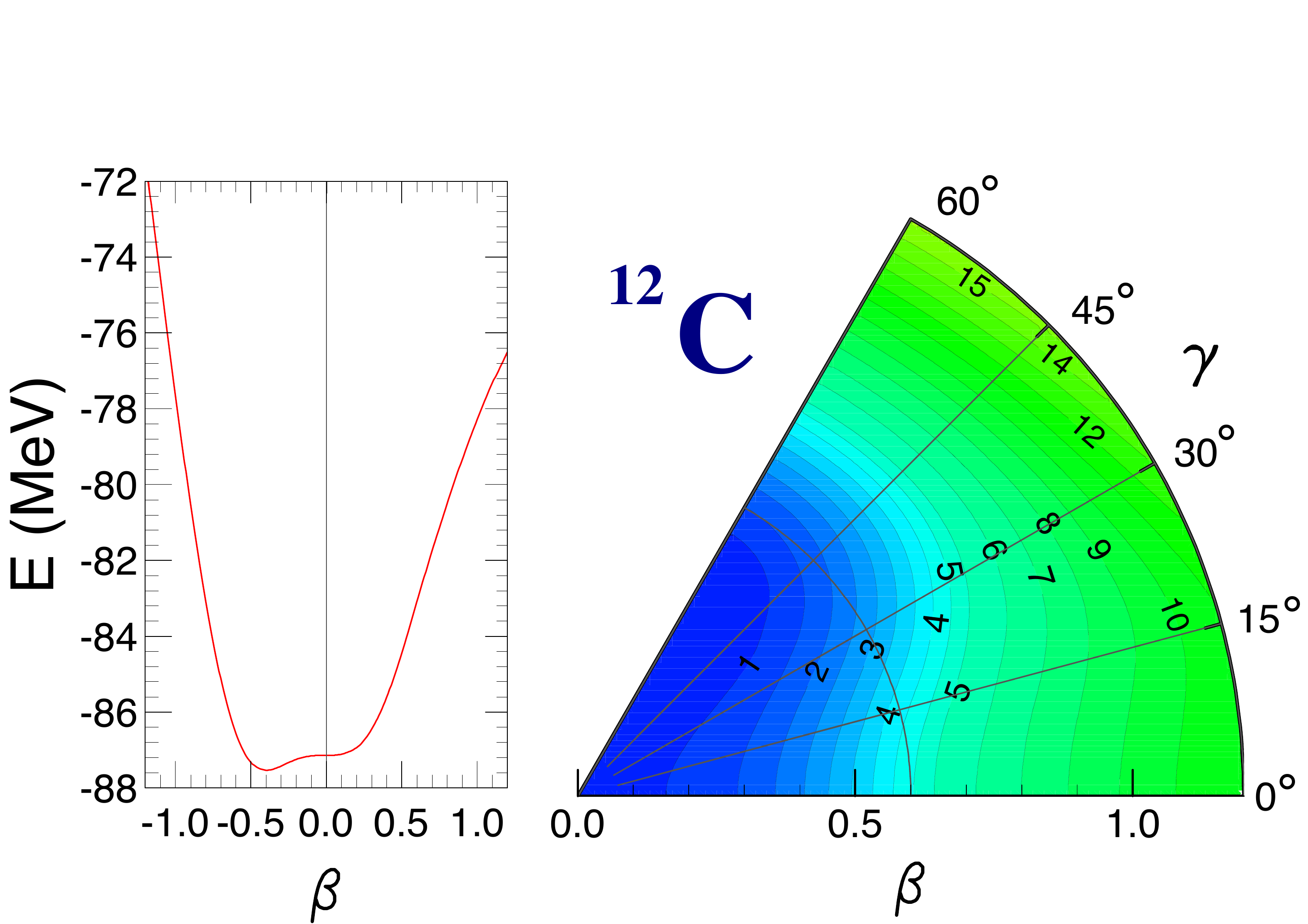}
\caption{(Color online) HFB axial Potential Energy Curve (left) and triaxial Potential Energy Surface (right) for the $^{12}$C nucleus.}
\label{f:12Ctriax}
\end{center}
\end{figure}

The iterative procedure discussed in the previous section is now applied to a test nucleus: $^{12}$C. At the mean-field level, this nucleus is predicted to be deformed and soft. For information we show in Fig.~\ref{f:12Ctriax} the potential energy curve (PEC) and potential energy surface (PES) provided by triaxial Hartree-Fock-Bogoliubov (HFB) calculations performed with the D1S Gogny force. Two minima appear. In particular the ground-state exhibits an oblate shape characterized by an axial deformation parameter $\beta\sim -0.40$. Although spherical symmetry is explicitly preserved in our approach (i.e. calculations are performed at $\beta \sim 0.0$ inducing states $\vert \Psi \rangle$ of good angular momentum J), the deformation properties of this nucleus should reflect in the correlation matrix $\sigma$.

\begin{figure}[t]
\centering
\begin{subfigure}[b]{0.25\textwidth}
\includegraphics[width=\textwidth]{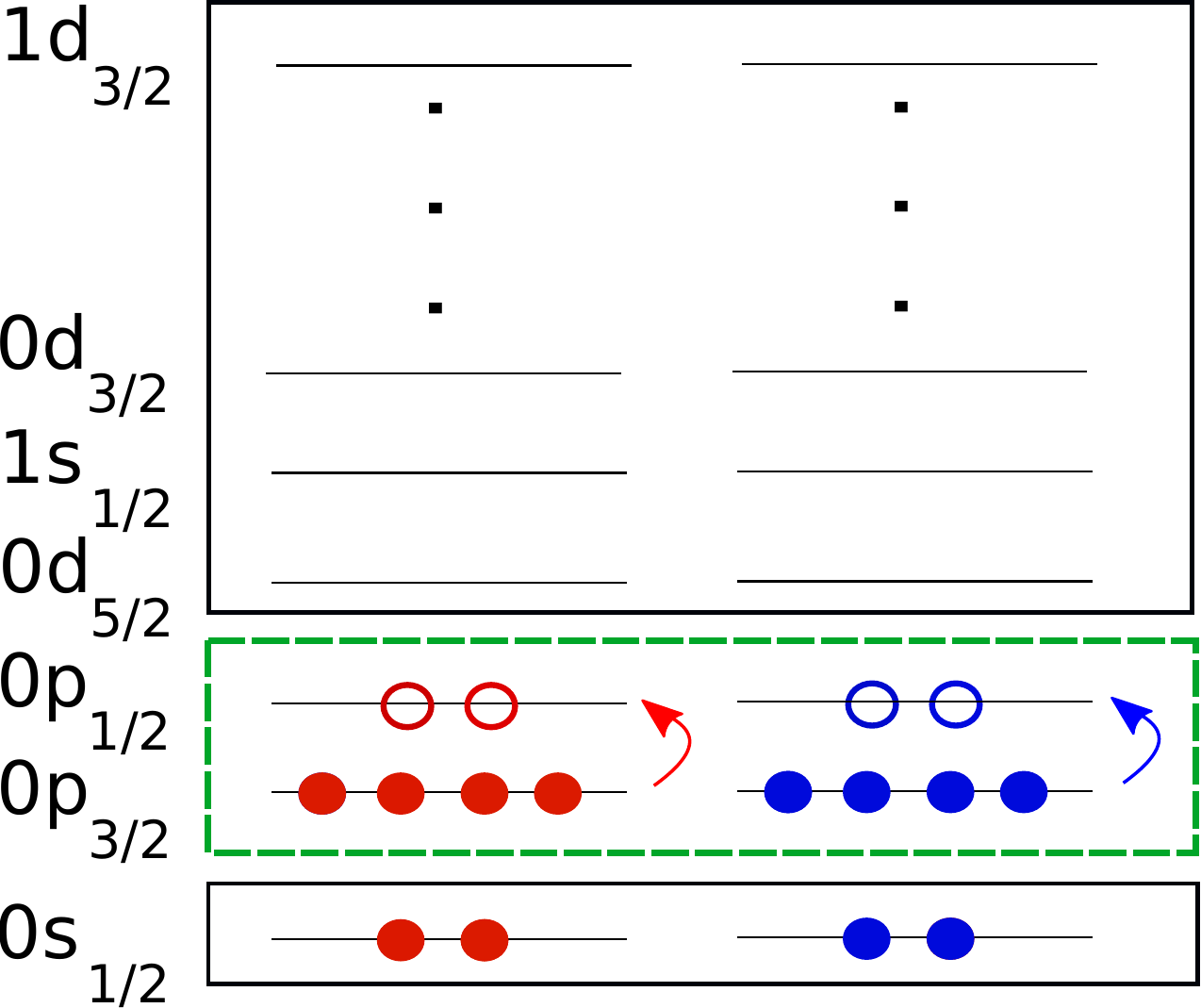}
\caption{Scheme 1}
\label{f:fig_trunc1}
\end{subfigure}
\begin{subfigure}[b]{0.209\textwidth}
\includegraphics[width=\textwidth]{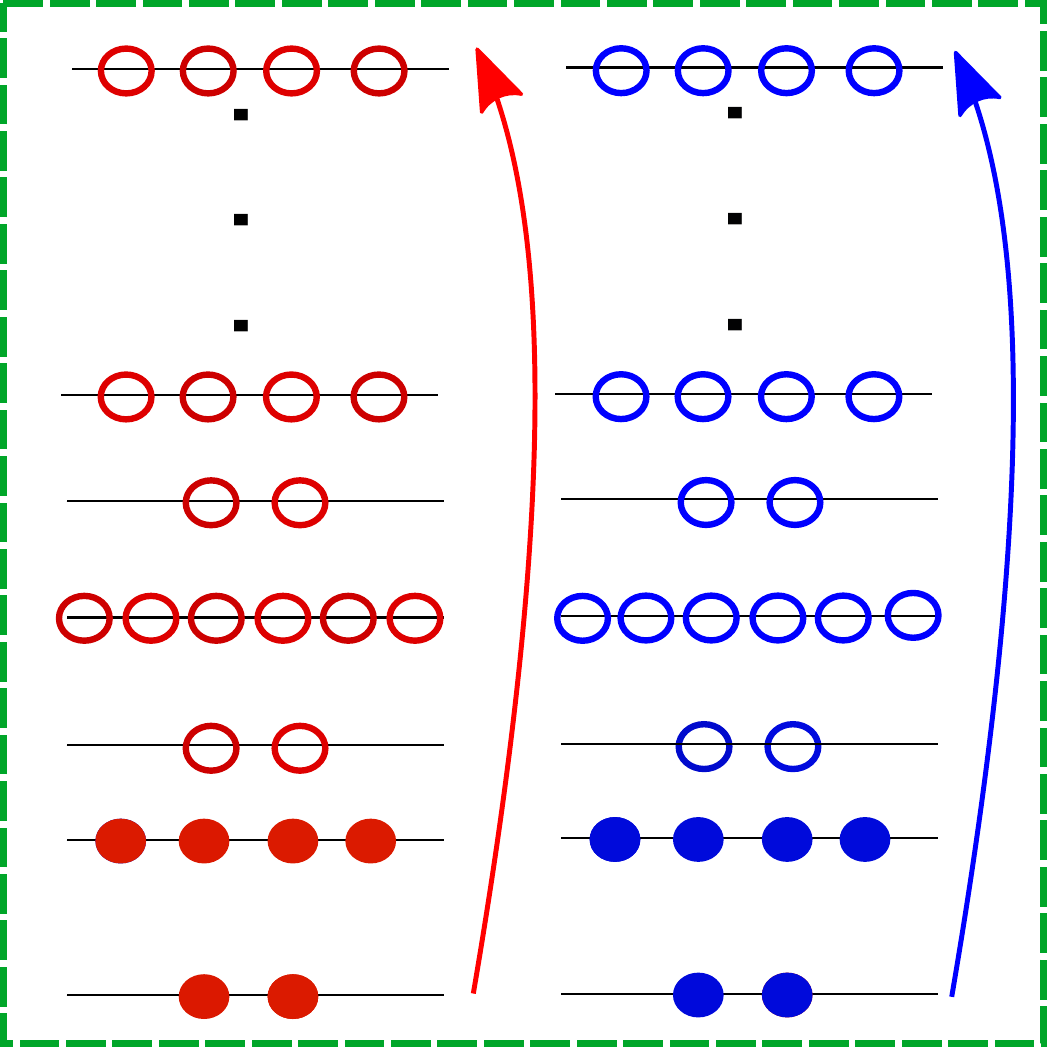}
\caption{Scheme 2}
\label{f:fig_trunc2}
\end{subfigure}
\caption{(Color online) Schematic representation of the two truncation schemes.}
\label{f:fig_trunc}
\end{figure}

\noindent The small number of particles contained in the $^{12}$C nucleus allows to test different types of truncation schemes of the wave function. 
Here we will compare:
\begin{enumerate}
\item \label{trunc1} A Shell-Model type truncation, dividing the single-particle space into three blocks: a filled core of $^{4}$He, a $0p$ valence shell, and remaining empty orbitals. In this scheme, all possible nucleon excitations in the valence space are explicitly introduced in the configuration mixing: 
$\ket{\Psi}=\sum_{\alpha \in model space} A_\alpha \ket{\phi_\alpha}$. In this way, excitations up to four particles - four holes ($4p$-$4h$) 
can be generated. The maximum value of the excitation order for both protons and neutrons, is $M_{\alpha_\pi}=M_{\alpha_\nu}=2$. This truncation scheme is represented in Fig.~\ref{f:fig_trunc1}. In the following we will see how the blocks of orbitals (core, valence space and empty states) are being mixed through the transformation of single-particle states, and thus how none of these initial spaces remain frozen.
\item \label{trunc2} A truncation based on the excitation order of the many-body configurations. Here we include all possible proton and neutron excitations up to $2p$-$2h$ in the full single-particle space. This generates A-body states up to $4p$-$4h$. In this way, we are able to quantify the full effect beyond the restricted valence space. No use of a core is made in this scheme, so that all particles are explicitly active at all times. This is sketched in Fig.~\ref{f:fig_trunc2}.
\end{enumerate}

\noindent One of the goals of this illustrative study is to validate the algorithm described previously, as well as to appreciate and compare the effect of the orbital optimization according to the content of the wave function. Moreover, this primordial study will provide us with new information on the D1S Gogny interaction when it is used within such kind formalism, with various truncation schemes.

\subsection{Numerical details}

In the present calculations, the single-particle states are expanded on axially deformed harmonic oscillator states, so that the many-body configurations are explicitly characterized by a good projection $K \equiv J_z$ of the angular momentum $J$ (the so-called m-scheme). The calculations are done at the spherical point ($\beta=0$), that is, the perpendicular and longitudinal oscillator frequencies are taken equal: $\omega_{\bot}=\omega_z \equiv \omega$. In this way, the two truncation schemes employed here generate rotationally invariant many-body spaces and thus produce a correlated state $\ket{\Psi}$ with a definite value of $J$. The self-consistent property of the spherical symmetry ensures its preservation along the convergence process. This feature allowed us to check the accuracy of our numerical code. The values of the oscillator frequency $\omega$, as well as the number of major shells $N_0$ are optimized at the Hartree-Fock level, leading to $\hbar \omega =15.50$ and $N_0=5$ shells. Sections \ref{subs_correl} to \ref{subs_groundstate} being dedicated to the study of the ground state of the even-even nucleus $^{12}$C, we have $J=K=0$. The configurations $\ket{\phi_{\alpha}} = \ket{\phi_{\alpha_{\pi}}} \otimes \ket{\phi_{\alpha_{\nu}}}$ are classified into blocks of projections ($K_{\alpha_{\pi}},K_{\alpha_{\nu}}=K-K_{\alpha_{\pi}}$) and organized by increasing excitation orders ($0p$-$0h$, $1p$-$1h$, $2p$-$2h$...). Time reversal invariance allows to deduce the configuration blocks with ($K_{\alpha_{\pi}}>0$) from the ones characterized by ($K_{\alpha_{\pi}}<0$). The former are therefore never explicitly built and the size of the matrix $\mathcal{H}[\rho,\sigma]$ to diagonalize is drastically reduced (by a factor $\sim 2$). 
\\ \\
With these conventions, the number of configurations obtained with the truncation scheme \ref{trunc1} is equal to 38, whereas it reaches 26 401 700 when using the truncation scheme \ref{trunc2}. Global convergence with the criteria on the one-body density mentioned in the previous section is reached in 15 and 14 macro-iterations when using the schemes 1 and 2 respectively.
\\ \\
Finally, we note that the center-of-mass motion is only corrected at the one-body level, i.e. the kinetic energy in Eq. ( \ref{e:GognyHamilt}) is multiplied by the factor $(1-\frac{1}{A})$. Although two-body corrections should also be taken into account, they are not implemented in this work.

\subsection{Evolution of correlations and densities along the convergence process} \label{subs_correl}

In this next section, we discuss the results obtained with the schemes 1 and 2. They will be represented on the various figures in green and blue respectively (online).

\subsubsection{Two-body correlation matrix $\sigma$} 
\begin{figure*}[t]
\centering
\begin{subfigure}[b]{0.49\textwidth}
\includegraphics[width=\textwidth]{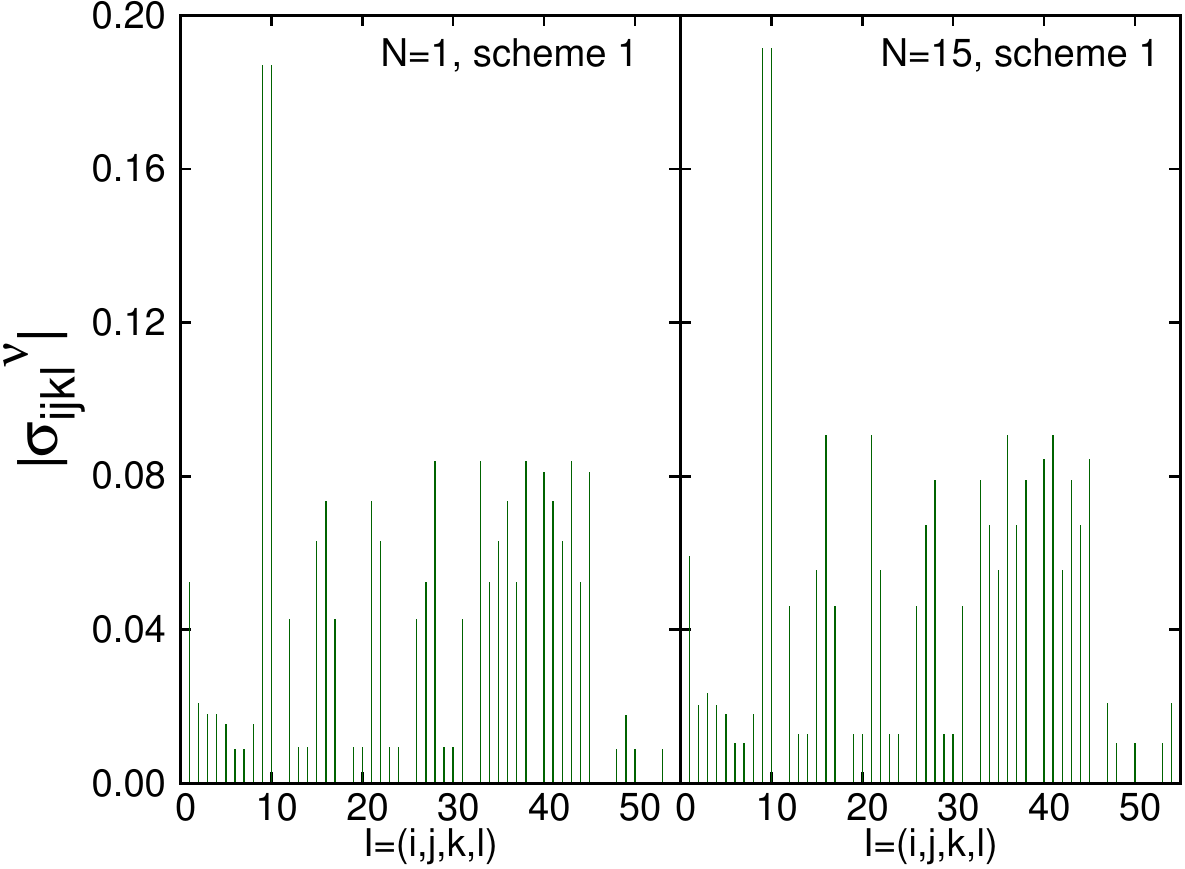}
\caption{Neutron correlations obtained with scheme 1 at macro-iteration N=1 (left) and N=15 (right).}
\label{f:sigma_neut_t1}
\end{subfigure}
\begin{subfigure}[b]{0.49\textwidth}
\includegraphics[width=\textwidth]{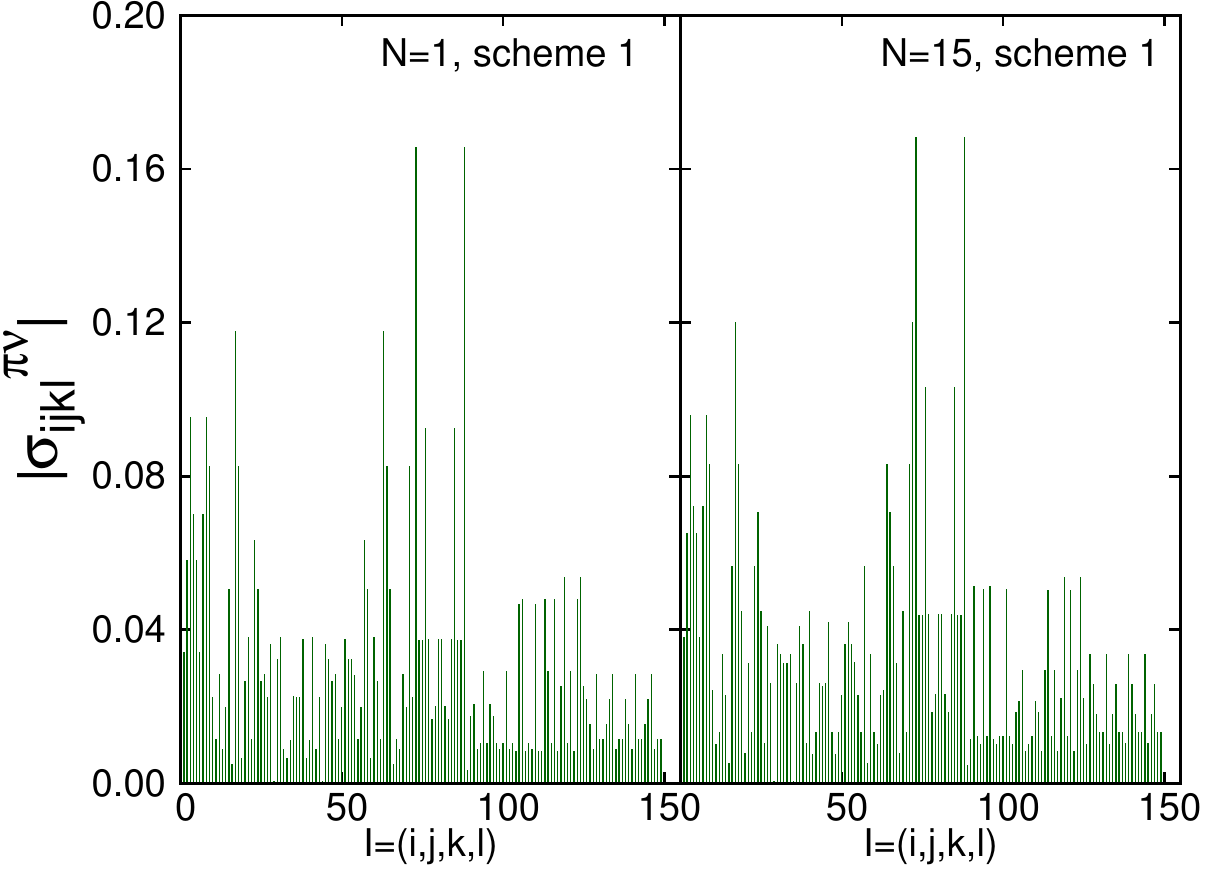}
\caption{Proton-Neutron correlations obtained with scheme 1 at macro-iteration N=1 (left) and N=15 (right).}
\label{f:sigma_protneut_t1}
\end{subfigure}
\begin{subfigure}[b]{0.49\textwidth}
\includegraphics[width=\textwidth]{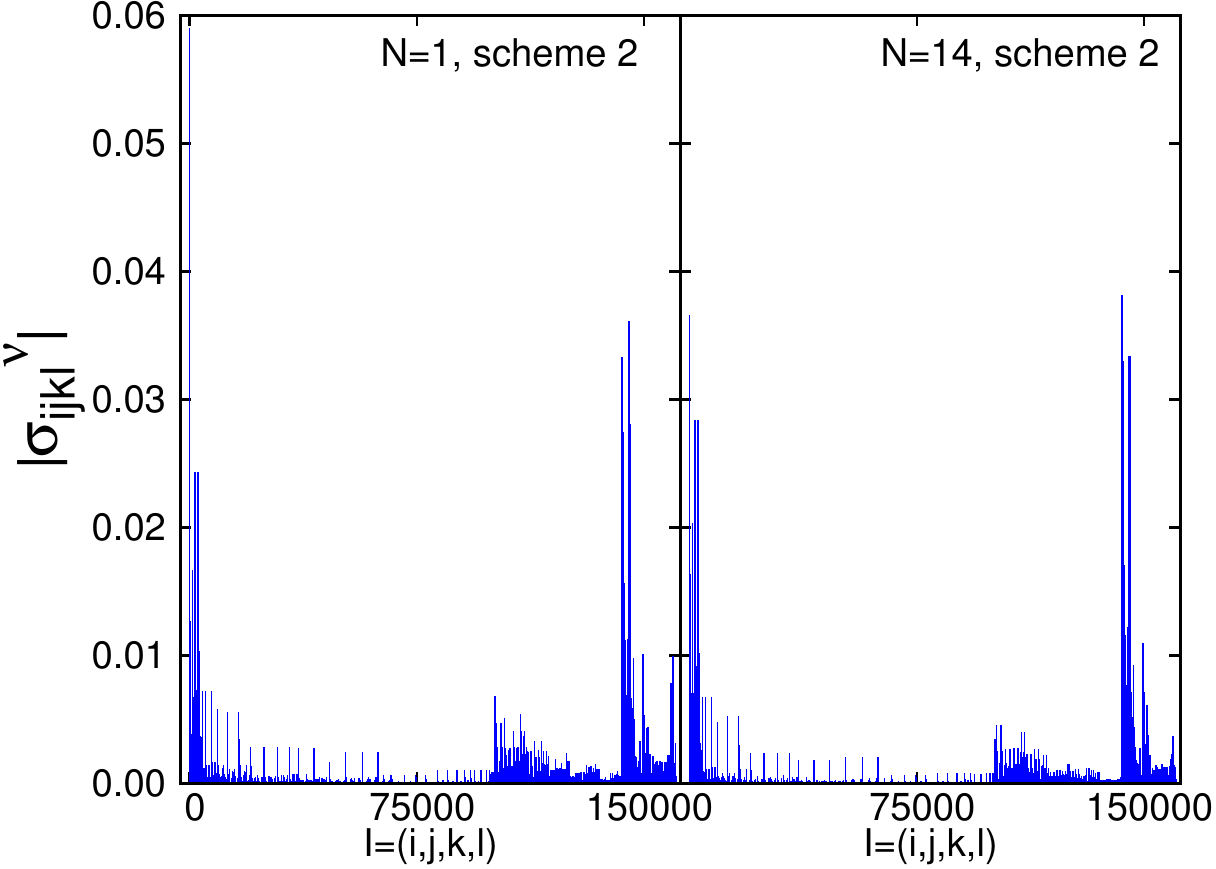}
\caption{Neutron correlations obtained with scheme 2 at macro-iteration N=1 (left) and N=14 (right).}
\label{f:sigma_neut_t2}
\end{subfigure}
\begin{subfigure}[b]{0.49\textwidth}
\includegraphics[width=\textwidth]{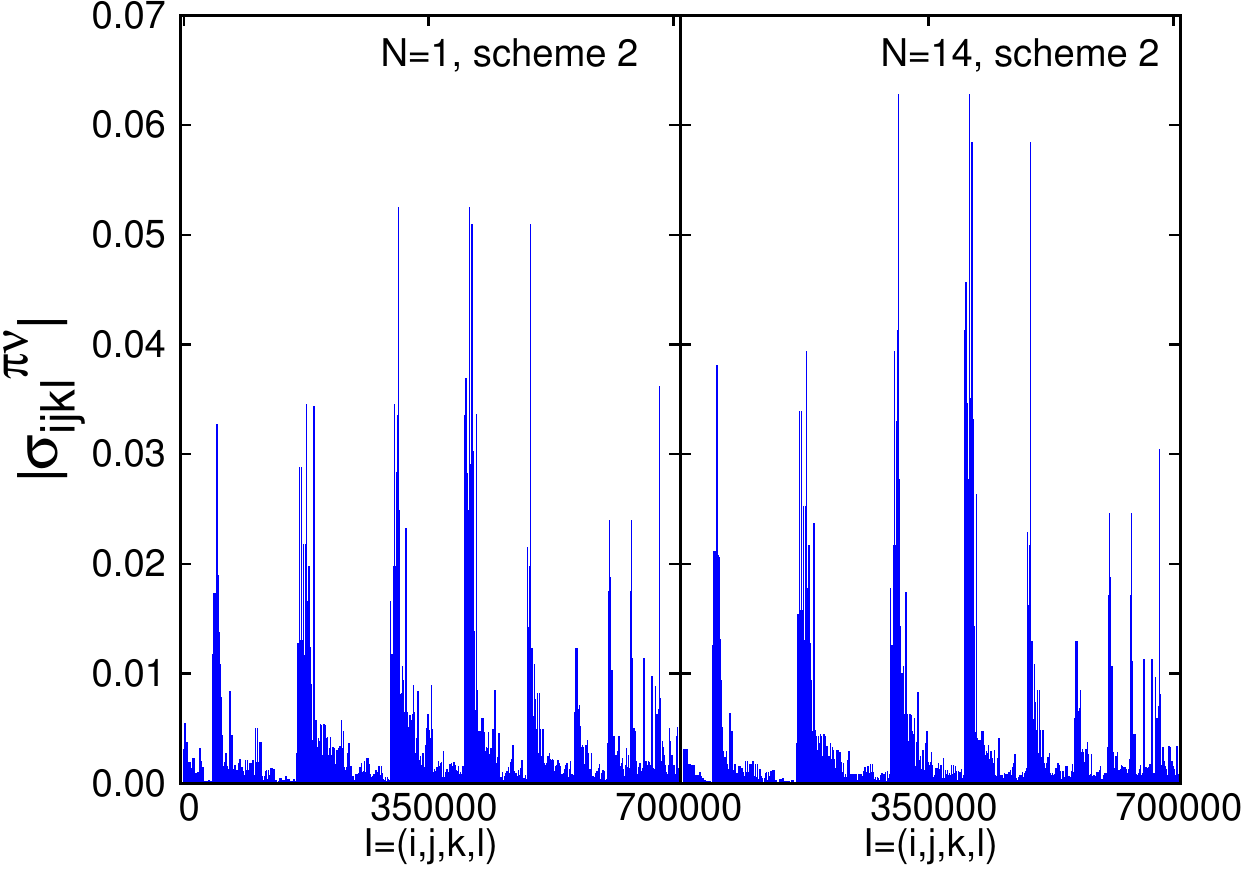}
\caption{Proton-Neutron correlations obtained with scheme 2 at macro-iteration N=1 (left) and N=14 (right).}
\label{f:sigma_protneut_t2}
\end{subfigure}
\caption{(Color online) Correlation matrices.}
\end{figure*}

The first step of the method is to diagonalize the many-body matrix $\mathcal{H}[\rho^{(0)},\sigma^{(0)}]$ to obtain the ground state expansion coefficients $\{A_\alpha^{(1)}\}$. From these we can calculate the two-body correlation matrix $\sigma^{(1)}$. We show on the left-hand side of Figs.~\ref{f:sigma_neut_t1} and \ref{f:sigma_neut_t2} the neutron correlation matrices calculated at macro-iteration $N=1$ with the truncation scheme \ref{trunc1} and \ref{trunc2} respectively. Since $^{12}$C has $N=Z$, the results for protons are very similar to the one obtained for neutrons and thus are not shown here. The proton-neutron correlation matrices calculated at $N=1$ with schemes 1 and 2 are shown on the left-hand side of Figs.~\ref{f:sigma_protneut_t1} and \ref{f:sigma_protneut_t2} respectively. All non-zero elements $|\sigma_{ijkl}|$ are plotted in absolute value, in order to appreciate the strength of correlations in each case. The horizontal axis is a linear index $I$ corresponding to a certain quadruplet of single-particle indices $I\equiv(i,j,k,l)$. Let us note that these correlation matrices are not recoupled in total angular momentum, it is therefore difficult to compare the intensity of the couplings between different shells and only qualitative remarks can be made here.

Comparing the two truncation schemes, we clearly note the higher fragmentation of the correlation matrices when the full single-particle space is active. Many more matrix elements of smaller intensity appear, which is expected to reveal a higher collectivity. About $50$ non-zero elements of the neutron (or proton) correlation matrix can be calculated with a p-shell valence space, whereas they amount to more than $150\,000$ when all orbitals are active. In both cases, correlations between protons and neutrons seem to dominate in number and intensity. One obtains about $150$ non-zero elements with the truncation scheme 1 and more than $700\,000$ with the truncation scheme 2. This type of correlations are usually very important to explain the deformation mechanism in nuclei. Moreover, in $N=Z$ nuclei, as both types of nucleons occupy the same orbitals the T=0 residual interaction is expected to be enhanced. We note that in all cases, the peaks with the highest magnitude correspond to correlations of pairing type, reflecting the scattering of pairs of particles around the Fermi level.
\\ \\
We discuss now in a qualitative way the intensity of the correlation matrices calculated at the end of the convergence procedure. We show on the right-hand side of Figs.~\ref{f:sigma_neut_t1} and \ref{f:sigma_neut_t2} the neutron correlations obtained when global convergence has been reached, for the scheme 1 and 2 respectively. In order to appreciate to what extent the new basis has "absorbed" part of the correlations, the quadruplet I now represent states $(i,j,k,l)$ of the optimized self-consistent single-particle basis. Similarly the proton-neutron correlations obtained after convergence are shown on the right-hand side of Figs.~\ref{f:sigma_protneut_t1} and \ref{f:sigma_protneut_t2}. Concerning the p-shell calculation (Figs.~\ref{f:sigma_neut_t1} and \ref{f:sigma_protneut_t1}), the correlation content does not seem much modified 
by the optimization of mean-field and orbitals. In fact we find that the sum of all non-zero elements increases slightly from $\sum |\sigma_{ij,kl}^{\nu}|=2.11$ at iteration $N=1$, to $2.27$ at iteration $N=15$ for the neutron correlations, and from $\sum |\sigma_{ij,kl}^{\pi\nu}|=4.58$ to $5.02$ in the case of proton-neutron correlations. A possible explanation could be that not enough information was explicitly introduced in the wave function, so that the correlation content is not important enough for the orbital equation to respond. When the full single-particle space is active (Figs.~\ref{f:sigma_neut_t2} and \ref{f:sigma_protneut_t2}), we note a decrease of some elements of the correlation matrices of same isospin (e.g. peak at I=6). In fact the sum $\sum |\sigma_{ij,kl}^{\nu}|$ of all elements of neutron type goes from $17.81$ at iteration $N=1$ to $14.68$ at $N=14$. At first glance, correlations of proton-neutron type seem to have increased after the convergence procedure. However we find that the sum $\sum |\sigma_{ij,kl}^{\pi\nu}|$ also decreases from $89.57$ to $74.40$. This behavior is coherent with the interpretation of the role of the orbital equation. The mean-field is indeed supposed to absorb as much effect of the correlations as possible and thus reduce the intensity of the latter. Although decreased we note that the residual correlations remain important. Let us remind that the spherical symmetry is explicitly preserved in our approach, and it is likely that more correlations could be absorbed by the mean-field if one allowed for deformation. Hence, it would be very informative to perform the same study by working in the intrinsic frame of the nucleus. The main drawback of such an approach would however be the need to project the final solution in order to obtain a state characterized by a good angular momentum $J$.

\subsubsection{Source term}
\begin{figure}[h]
\begin{center}
\includegraphics[width=\columnwidth]{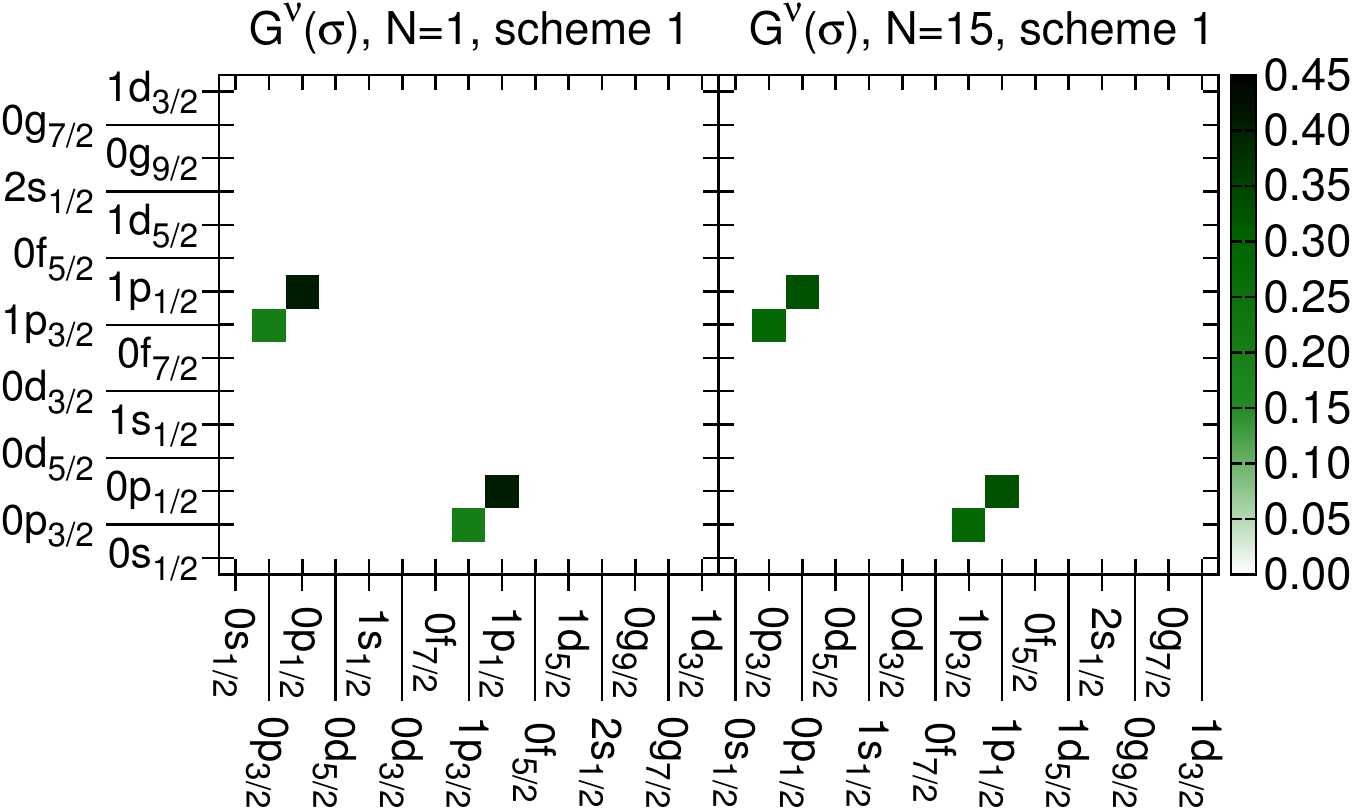}
\caption{(Color online) Absolute value of the neutron source term in MeV, obtained with scheme 1 at macro-iteration N=1 (left) and N=15 (right).}
\label{f:gsig_T1}
\end{center}
\end{figure}

The previous correlation matrices are now used to calculate the source term $G[\sigma]$ appearing in the orbital equation (\ref{e:eq2_Gogny}). Let us first look more closely at the analytical expression of this term:
\begin{eqnarray}
G[\sigma]_{ij} &=&  \frac{1}{2} \sum_{klm} \sigma_{ki,lm} \braket{kl|\widetilde{V}^{2N}[\rho]|jm}   \nonumber \\  
               && \;\;\; -    \frac{1}{2}\sum_{klm} \braket{ik|\widetilde{V}^{2N}[\rho]|lm}  \sigma_{jl,km}   \; .
\end{eqnarray}
We note that $G[\sigma]_{ij}$ is non-zero if there exists at least one triplet $(k,l,m)$ of single-particle states such that $\sigma_{ki,lm} \ne 0$ or $\sigma_{jl,km} \ne 0$. Since $\sigma$ reflects the correlations that have been explicitly introduced in the wave function, $\sigma_{ki,lm} \ne 0$ if $(k,i,l,m)$ are all explicitly active. When the use of a valence space is made, the source matrix $G[\sigma]_{ij}$ therefore has at least one external index ($i$ or $j$) in this space. The second index being attached to the matrix element of the interaction $\widetilde{V}^{2N}[\rho]$, it can belong to the whole single-particle basis. The source term is therefore able to couple the active shell to the rest of the orbitals that were previously considered as inert. Thus, it has the role of propagating the effect of correlations on the full single-particle basis by establishing a communication between the three blocks (core/valence/empty single-particle states). Because of the explicit symmetry preservations imposed in this study, the source term can only couple states of same parity $\pi$ and angular momentum $j$. When using the truncation scheme 1, it therefore couples the $0p_{3/2}$ and $0p_{1/2}$ sub-shells to the $1p_{3/2}$ and $1p_{1/2}$ sub-shells respectively. This is illustrated in Fig.~\ref{f:gsig_T1} which represents the neutron source term at the beginning and end of the convergence procedure. More precisely, since states $i=(\alpha_i,m_i)$ of the same spherical subshell $\alpha_i=(n_i,l_i,j_i)$ lead to identical couplings $|G^{\nu}[\sigma]_{i,j}|=|G^{\nu}[\sigma]_{\alpha_i,\alpha_j}| \delta_{j_i,j_j} \delta_{m_i,m_j}$, we only plotted the values $|G^{\nu}[\sigma]_{\alpha_i,\alpha_j}|$ for the different subshells. At iteration $N=1$, the basis $\{i\}$ denotes the Hartree-Fock states, whereas it denotes optimized states at iteration $N=15$.

\begin{figure}[h]
\begin{center}
\includegraphics[width=\columnwidth]{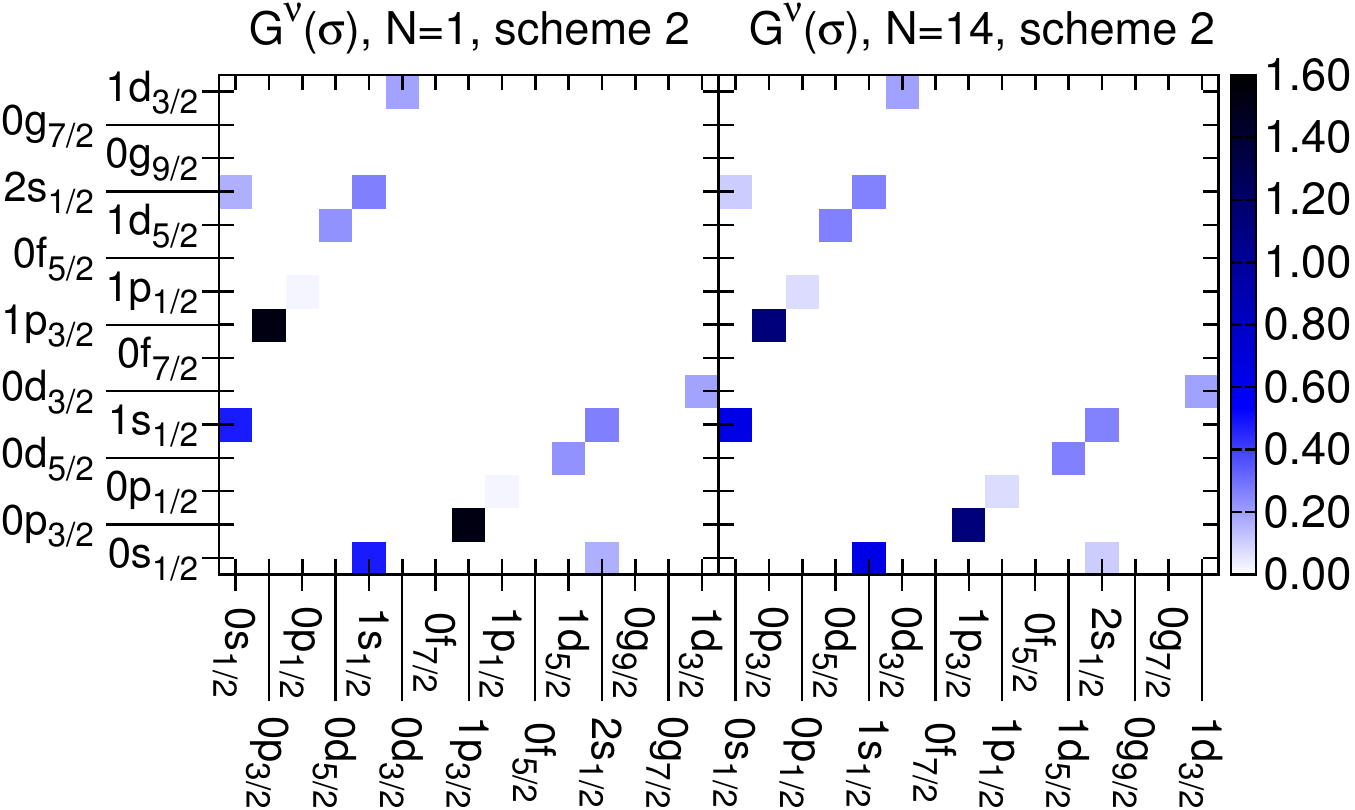}
\caption{(Color online) Absolute value of the neutron source term in MeV, obtained with scheme 2 at macro-iteration N=1 (left) and N=14 (right).}
\label{f:gsig_T2}
\end{center}
\end{figure}

\noindent Similarly, we show in Fig.~\ref{f:gsig_T2} the source term calculated with the scheme 2. Since all orbitals participate to the configuration mixing, many more couplings appear in the source matrix $G_{ij}$, than in the previous case, where a restricted valence space was considered. Also, the values of the couplings between the $0p$ and $1p$ shells have drastically changed: the coupling between the states of the $p_{3/2}$ sub-shells is now equal to $\sim 1.44$ MeV while it was only $\sim 0.21$ MeV with the truncation scheme 1. On the contrary the coupling between the orbitals of the $p_{1/2}$ sub-shells has been decreased from $\sim 0.42$ to $\sim 4.8\times 10^{-3}$ MeV. \\

\noindent Let us now comment on the evolution of $G[\sigma]$ along the convergence procedure. This source term  reflects the residual correlations beyond the mean field $h[\rho,\sigma]$. Since the latter absorbs the average effect of the correlation content of the system and thus becomes more and more refined, one could expect the intensity of $G[\sigma]$ to decrease. However, as seen from the right-hand side of Figs.~\ref{f:gsig_T1} and \ref{f:gsig_T2}, the results obtained are not so straightforward. In both cases, some kind of "harmonization" of the different couplings seems to appear: the strongest ones decrease while the weakest ones increase. Concerning the scheme 1, the sum $\sum_{\alpha_i<\alpha_j} |G^\nu_{\alpha_i\alpha_j}[\sigma]|$ varies from $0.669$ MeV at iteration $N=1$, to $0.663$ MeV at iteration $N=15$, while the average value of the couplings varies from $0.335$ to $0.332$ MeV. The effect is slightly more important when using the scheme 2: the sum of the couplings decreases from $2.809$ to $2.634$ MeV, while the average evolves from $0.401$ to $0.376$ MeV.

\subsubsection{One-body density}

\begin{figure}[t]
\begin{center}
\includegraphics[width=\columnwidth]{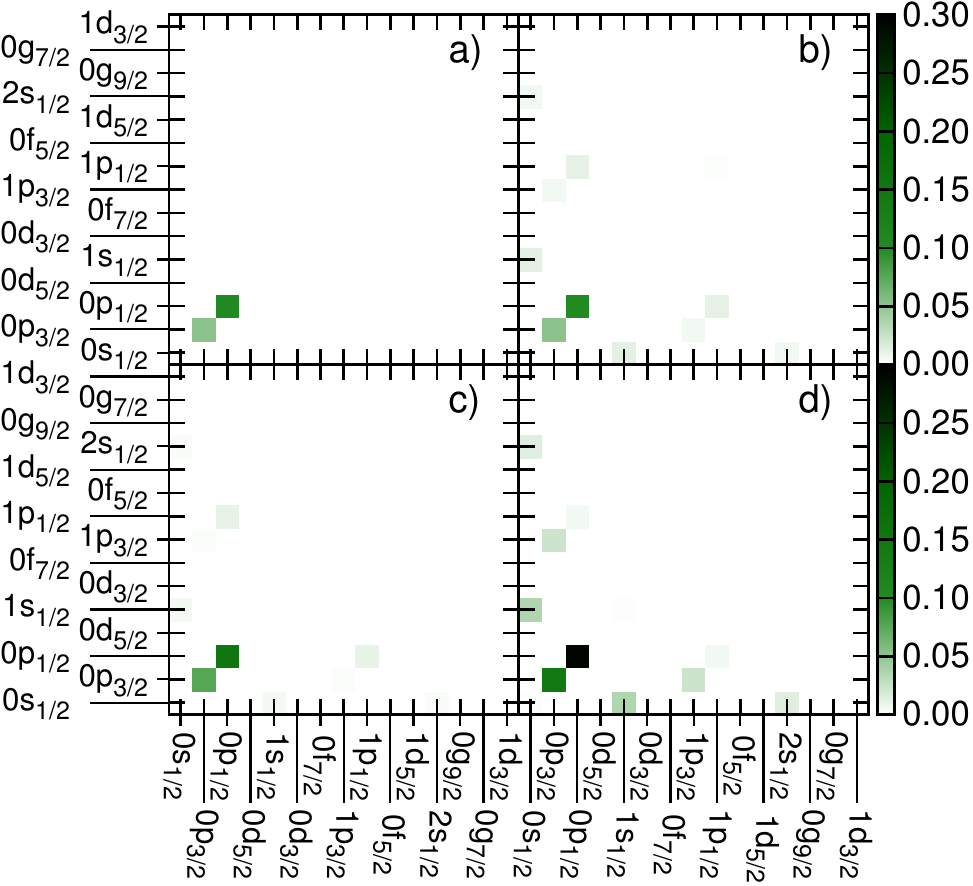}
\caption{(Color online) Evolution of the neutron one-body density along the convergence process. The difference between the correlated density and a pure Hartree-Fock density $\Delta \rho = |\rho - \rho_{HF}|$ is represented in a matrix form, in the original Hartree-Fock basis. Truncation scheme 1. See text for explanation.}
\label{f:rho_evol1}
\end{center}
\end{figure}

\begin{figure}[t]
\begin{center}
\includegraphics[width=\columnwidth]{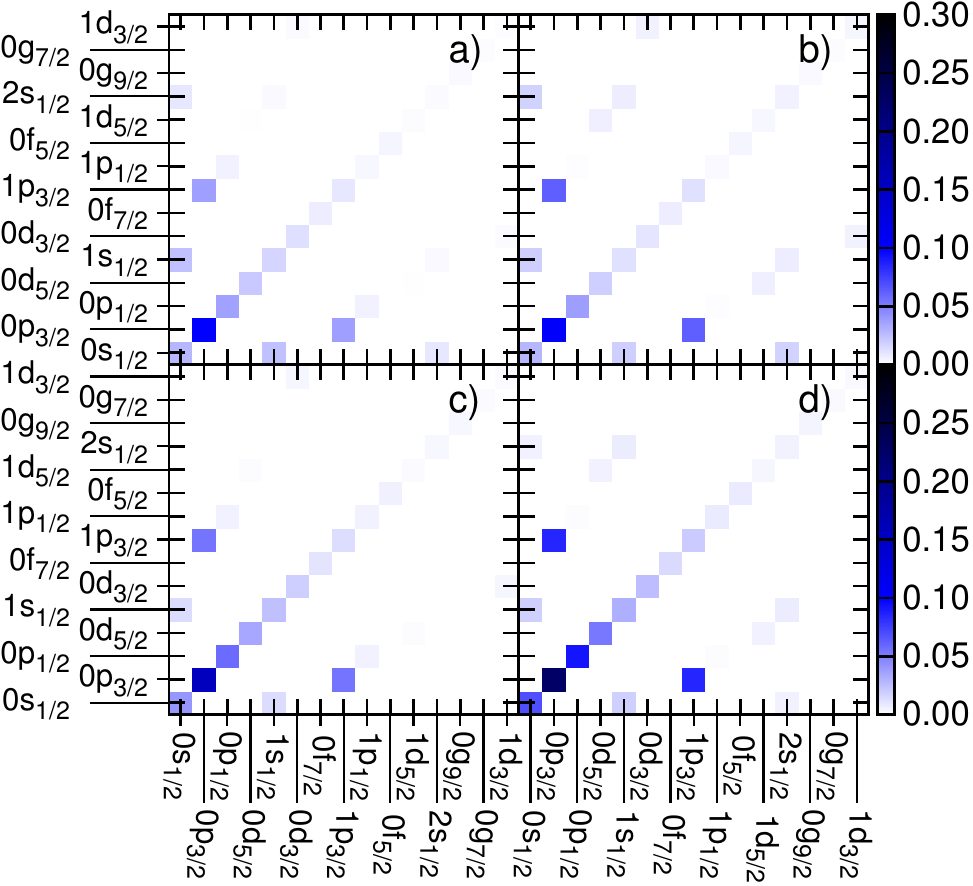}
\caption{(Color online) Evolution of the neutron one-body density along the convergence process. The difference between the correlated density and a pure Hartree-Fock density $\Delta \rho = |\rho - \rho_{HF}|$ is represented in a matrix form, in the original Hartree-Fock basis. Truncation scheme 2. See text for explanation.}
\label{f:rho_evol2}
\end{center}
\end{figure}

A first manifestation of the effect induced by the orbital equation can be seen on the evolution of the one-body density matrix $\rho$. Before looking at the results, it is important to remind the following. In the previous section, we mentioned the density $\rho^{init}$ calculated from the output of the first variational equation (\ref{e:eq1_Gogny}), i.e. calculated as
\[ \rho^{init}_{ij} = \sum_{\alpha\beta} A^*_{\alpha} A_{\beta} \braket{\phi_{\alpha}|a^{\dagger}_j a_i|\phi_{\beta}} \; , \]
and the density resulting from the second variational equation (\ref{e:eq2_Gogny}), i.e. solution of $[h,\rho]=G$. Formally these two densities should correspond to the same quantity. However, at the beginning of the procedure, when convergence has not been yet reached, they are not identical. This is illustrated in Figs.~\ref{f:rho_evol1} and \ref{f:rho_evol2}, where we show the evolution of the neutron density along the convergence process, for the truncation schemes 1 and 2, respectively. Again, since $N=Z$ the proton density shows a similar behavior. To emphasize the effect of the orbital equation, we plotted the difference between the correlated density and the density of a pure Hartree-Fock state, in the original Hartree-Fock basis: $\Delta \rho_{\alpha_i,\alpha_j} \equiv |\rho_{\alpha_i,\alpha_j}-\rho^{HF}_{\alpha_i,\alpha_j}|=|\rho_{ij}-\rho^{HF}_{ij}|$, $\forall \, m_i=m_j=-j_i,...,j_i$.
\\ \\

\begin{figure}[t]
\begin{center}
\includegraphics[width=\columnwidth]{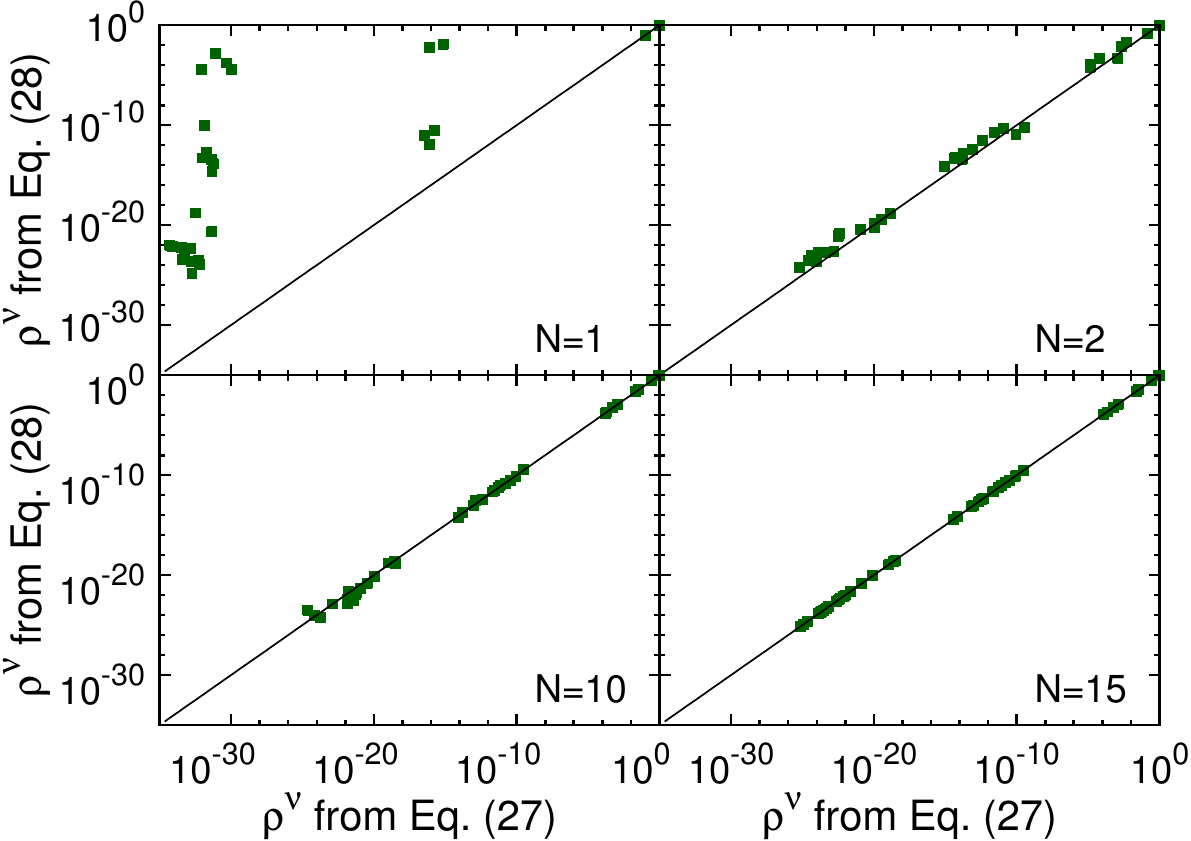}
\caption{(Color online) Comparison between the neutron density matrices given by the first and second variational equations (\ref{e:eq1_Gogny}) and (\ref{e:eq2_Gogny}) 
respectively, at different stages of the convergence process. Truncation scheme 1.}
\label{f:compareq12_trunc1}
\end{center}
\end{figure}

\noindent We start with the $0\hbar\Omega$ truncation scheme 1. In Fig.~\ref{f:rho_evol1}a we show the matrix $\Delta \rho$ obtained at the first macro-iteration $N=1$, resulting from the solution of the first equation (\ref{e:eq1_Gogny}) only (i.e. when the mixing coefficients have been calculated with the initial Hartree-Fock orbitals). As expected, the density is only modified in the valence space, where explicit correlations have been introduced. 
\\
\\
In Fig.~\ref{f:rho_evol1}b is represented $\Delta \rho$ obtained at $N=1$, after solving the orbital equation (\ref{e:eq2_Gogny}). We see that optimizing the single-particle states has modified the density in the whole basis and introduced non-diagonal elements $\rho_{\alpha_i\alpha_j}$. Couplings between positive-parity states also appear. Even though they have not been introduced in the configuration mixing, and thus are not affected by $G[\sigma]$, they are transformed through $\left[\hat{\mathpzc{h}}[\rho,\sigma] , \hat \rho\right]=0$ by the fact that the mean-field $\mathpzc{h}[\rho,\sigma]$ is much richer than a pure Hartree-Fock field, since it is polarized by the residual interaction. These states are indeed influenced by the two-body correlations in two ways: indirectly through the fact that the average potential is built with the correlated one-body density $\rho$, and directly through the rearrangement terms $\mathpzc{R}[\rho,\sigma]$ that introduce an explicit dependence on the two-body correlation matrix $\sigma$.
\\
\\
In Fig.~\ref{f:rho_evol1}c we show $\Delta \rho$ at the macro-iteration $N=2$ after solving the first variational equation (\ref{e:eq1_Gogny}). At this stage we redefined the $p$-shell valence space on the new single-particle basis. We note that the density kept trace of the orbital mixing and is starting to look similar to the density resulting from the orbital equation (\ref{e:eq2_Gogny}). In fact, as expected, we observe that the density matrices from both equations converge to the same quantity at the end of the procedure. This is shown in Fig.~\ref{f:compareq12_trunc1} where we plotted both densities at different stages of the convergence process. We see that they tend to align themselves on the $y=x$ line after a few iterations.
\\
\\
Finally we show in Fig.~\ref{f:rho_evol1}d the matrix $\Delta \rho$ obtained at the end of the convergence procedure (at macro-iteration $N=15$).
We see that the difference to the Hartree-Fock density has generally increased. Let us note that mixing the orbitals not only allows to introduce non-diagonal couplings in the density (which in this case would be nonexistent if only the first equation (\ref{e:eq1_Gogny}) was solved), it also modifies the diagonal elements of $\rho$. More precisely it allows in principle to partially empty the core and populate the initially empty states. In this test case, the biggest effect concerns the initial Hartree-Fock $0s$-shell of the core which is emptied by $~2.86\times 10^{-3}$ particles in the case of protons and $~2.96\times 10^{-3}$ in the case of neutrons. The initially empty $1s$-shell is populated by $~2.44\times 10^{-3}$ and $~2.54\times 10^{-3}$ particles respectively. In this test case, the effect is thus quite weak and not visible on the figures.
\\
\\

\begin{figure}[t]
\begin{center}
\includegraphics[width=\columnwidth]{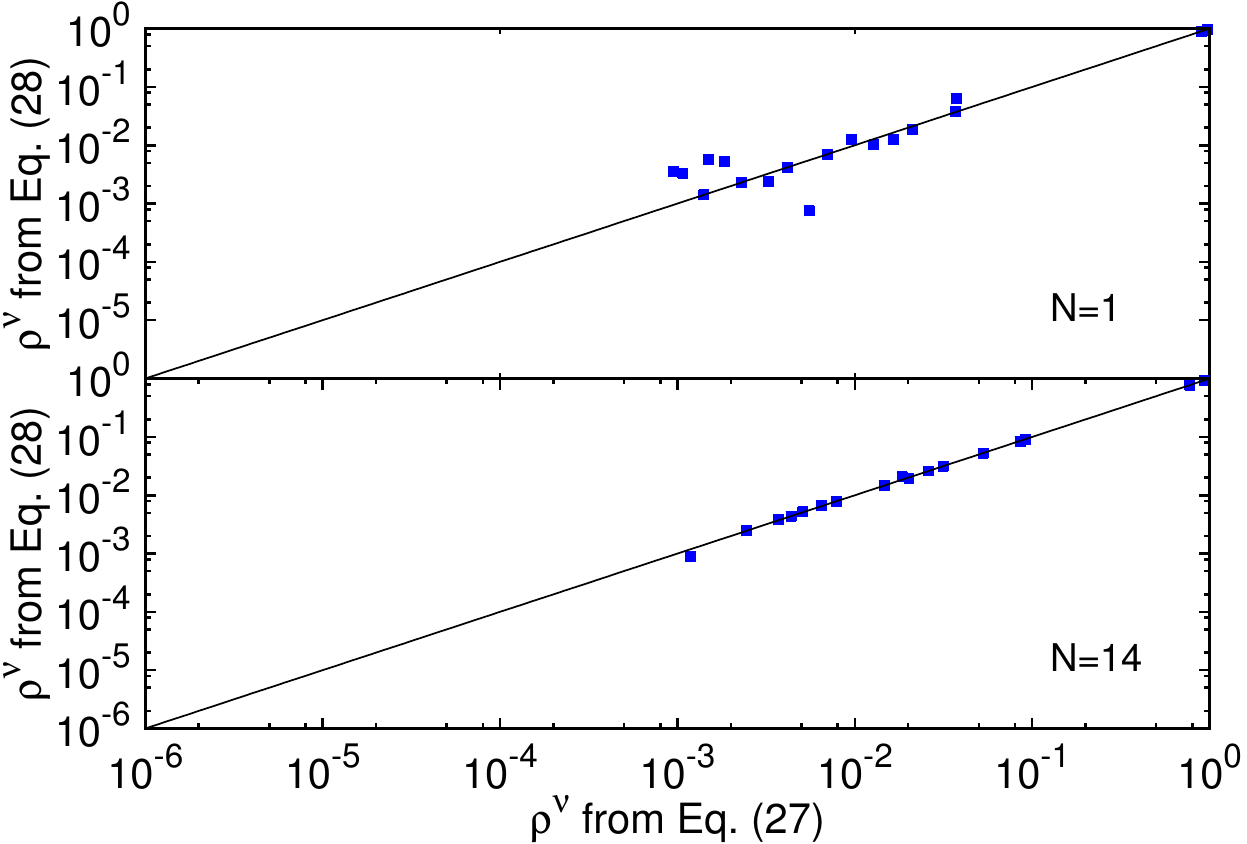}
\caption{(Color online) Comparison between the neutron density matrices given by the first and second variational equations (\ref{e:eq1_Gogny}) and (\ref{e:eq2_Gogny}) respectively, at different stages of the convergence process. Truncation scheme 2.}
 \label{f:compareq12_trunc2}
\end{center}
\end{figure}

\noindent Let us now look at the evolution of the density matrix obtained when all single-particle states are active (scheme 2). The results are shown in Fig.~\ref{f:rho_evol2} which is organized in the same way as Fig.~\ref{f:rho_evol1}. Since no use of a valence space is made, the correlated density calculated as output of the first variational equation (\ref{e:eq1_Gogny}) at macro-iteration $N=1$ already contains non-diagonal couplings. Still it is modified after solving the orbital equation (\ref{e:eq2_Gogny}). Again, the densities obtained via both variational equations tend to resemble each other along the convergence procedure. In fact they become identical up to $\sim 10^{-4}$, as seen from Fig.~\ref{f:compareq12_trunc2}. It is important to state that this also means that the non-diagonal elements of the density $\rho$ calculated via the first equation in the final optimal basis go to zero. Finally a stronger modification of the diagonal elements of the density is induced by the orbital optimization, compared to the previous truncation scheme. We show in Table~\ref{t:occup} the evolution of the "occupations" of the Hartree-Fock spherical subshells (diagonal elements of the density in the HF basis) at the beginning and end of the procedure. Identical behaviors are obtained for both protons and neutrons. We observe a depopulation of the $0s$ shell that is of the order of $0.08$ particles. More importantly $0.52$ particles leave the $0p_{3/2}$ subshell to populate higher shells.

\begin{table}[h]
 \centering
\begin{tabular}{ccc}
\hline
 Original Hartree-Fock       & Occupation at             & Occupation at             \\
 sub-shell                   & macro-iteration           & macro-iteration           \\
                             &$N=1$                      &$N=14$                     \\
\hline
\hline
$0s$                         & 1.94                      & 1.86                      \\
\hline
$0p_{3/2}$                   & 3.60                      & 3.08                      \\
\hline
$0p_{1/2}$                   & 0.074                     & 0.184                     \\
\hline
$0d_{5/2}$                   & 0.126                     & 0.318                     \\
\hline
\end{tabular}
\caption{Evolution of the "occupations" of the original Hartree-Fock spherical subshells for the truncation scheme 2. Results for neutrons.}
\label{t:occup}
\end{table}

\subsection{New single-particle energies}
\begin{figure}[h]
\begin{center}
\includegraphics[width=\columnwidth]{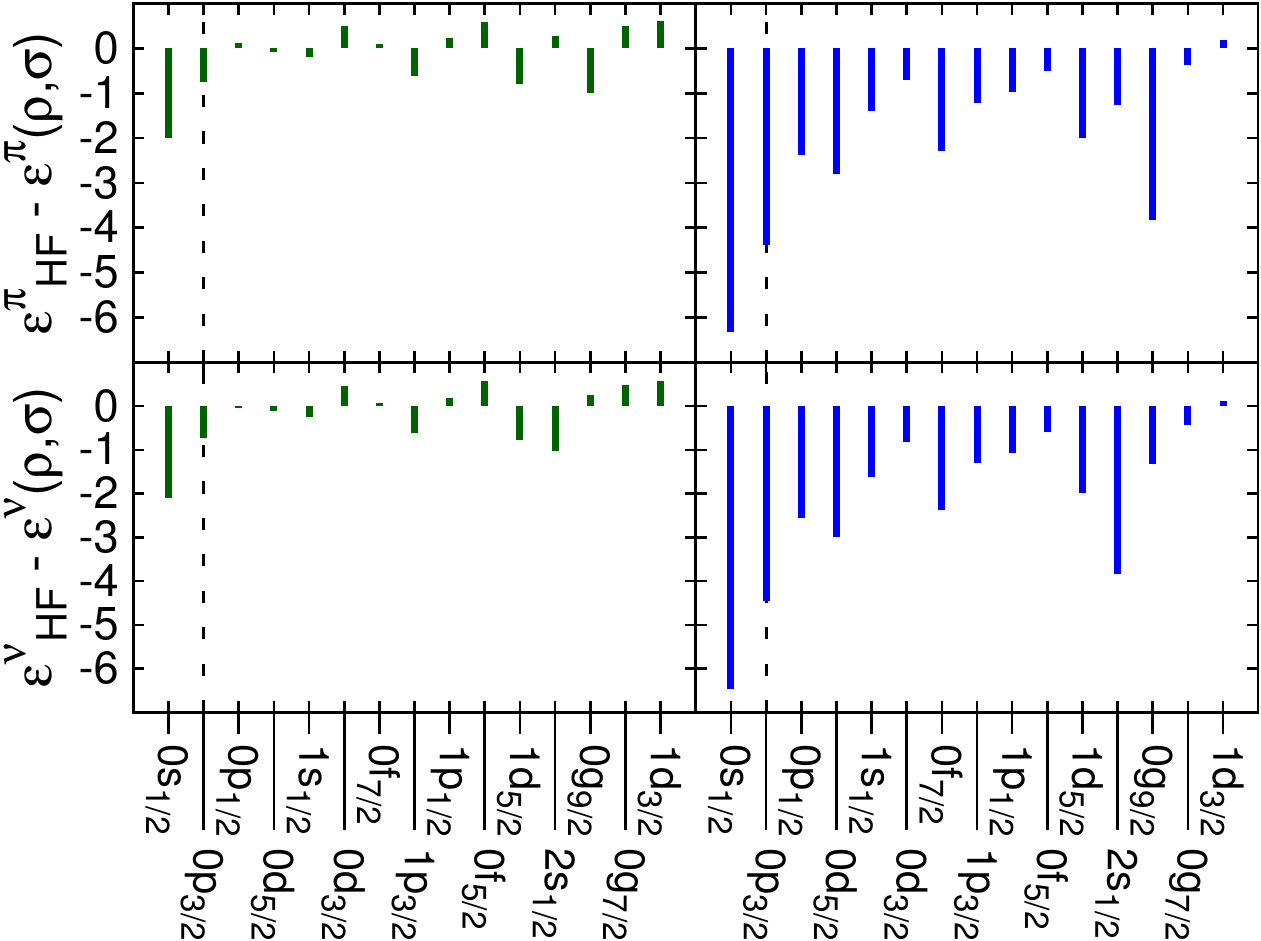}
\caption{(Color online) Difference $\Delta \varepsilon = \varepsilon_{HF} - \varepsilon[\rho,\sigma]$ in MeV, between single-particle energies taken as eigenvalues of the Hartree-Fock field and single-particle energies taken as eigenvalues of the improved mean-field $h[\rho,\sigma]$. Results obtained with truncation scheme 1 (left) and 2 (right) for protons (up) and neutrons (down). The Fermi levels are marked by a dashed line.}
\label{f:SPE}
\end{center}
\end{figure}

The single-particle energies (SPEs) $\varepsilon_\mu$ are defined as eigenvalues of the mean-field $h[\rho,\sigma]$. In order to appreciate the modification induced by the correlations on the single-particle spectrum, we plotted in Fig.~\ref{f:SPE} the difference between these SPEs and SPEs taken as eigenvalues of the pure Hartree-Fock field. Results obtained with the schemes 1 and 2 are shown on the left- and right-hand side of Fig.~\ref{f:SPE} respectively. Proton and neutron SPEs are on top and bottom of the figure respectively. 

\noindent We note that the account for correlations in the mean-field leads to a global compression of the single-particle spectrum in all cases. Similar results are obtained for protons and neutrons. The energy difference between the lowest shell ($0s$) and the highest one ($1d_{3/2}$ in this case) is decreased by $\sim 2.5$ MeV when using the truncation scheme 1 and by more than $6$ MeV with the truncation scheme 2. The most drastic effect concerns the $0s_{1/2}$-shell which is shifted up by more than $2$ MeV with the scheme 1 and $6$ MeV with the scheme 2. Moreover, the gap at the Fermi level (represented by a vertical dashed line) between the  $0p_{3/2}$ and  $0p_{1/2}$ sub-shells is reduced by $\sim 913$ KeV for protons and $\sim 740$ KeV for neutrons when using the truncation 1. It is reduced by $\sim 2$ MeV when using the truncation 2. These values are to be compared to the initial gap value of $\sim 8.15$ MeV in the Hartree-Fock calculation. Such a reduction of the gap is expected to favorize excitations around the Fermi level, and thus to induce modifications on the structure of the wave function. This is analyzed and discussed in the next section.

\subsection{Ground state properties} \label{subs_groundstate}
We now look at the effect caused by the orbital optimization on the energy and the composition of the ground state wave function. For a complete comparison and in order to isolate the effect of the orbital equation (\ref{e:eq2_Gogny}), we calculate these quantities at two levels:
\begin{itemize}
 \item After solving the first variational equation (\ref{e:eq1_Gogny}) with frozen Hartree-Fock orbitals, i.e. by iterating the diagonalization of the many-body matrix $\mathpzc{H}[\rho,\sigma] = H[\rho]+\mathpzc{R}[\rho,\sigma]$. 
 \item After the complete self-consistent procedure, when both variational equations (\ref{e:eq1_Gogny}) and (\ref{e:eq2_Gogny}) are simultaneously satisfied.
\end{itemize}

\subsubsection{Ground-state energy}

\noindent We first show in Table~\ref{t:Ecorr} the ground-state energies obtained in these two cases for the $^{12}$C nucleus.  
\begin{table}[h]
\begin{tabular}{ccc}
\hline
Scheme   & Eq.~(\ref{e:eq1_Gogny})  & Eqs.~(\ref{e:eq1_Gogny}) $\&$ (\ref{e:eq2_Gogny}) \\
\hline
\hline
  1        &    $-99.10$                &   $-99.44$                                          \\
\hline 
  2        &    $-154.65$               &   $-155.42$                                         \\
\hline
\end{tabular}
\caption{Ground-state energies in MeV.}
\label{t:Ecorr}
\end{table}

\noindent Using the D1S Gogny force, the energy of the spherical Hartree-Fock state is found to  be $E_{HF} \sim -92.88$ MeV, and already lies at the experimental value $E_{exp} \sim -92.16$ MeV \cite{Audi}. The additional correlations brought by the MPMH approach therefore inevitably lead to an overbinding of this nucleus. Using the first truncation scheme this overbinding is found to be reasonable with a value of $\sim 7$ MeV.
It can also be informative to compare the MPMH energies in Table~\ref{t:Ecorr} with the ones obtained in the HFB framework. Contrary to our case, the potential energy curve shown in Fig.~\ref{f:12Ctriax} has been obtained by including two-body center-of-mass corrections but without including the exchange term in the Coulomb field. Although the latter only modifies the energy by a few hundred keV, the inclusion of the two-body terms in the center-of-mass corrections can change the binding energy by several MeV. In this case, the axial HFB energy is brought down from $-87.54$ to $-92.45$ MeV when excluding these corrections and again lies at $\sim 7$ MeV above the MPMH result obtained with the first scheme. Let us stress that the  axial HFB total energy differs from the spherical HFB energy only by $\sim 400$ keV. This shows that the correlations added by MPMH go beyond the ones related to the axial deformation.
Finally, comparing the second and third columns of Table~\ref{t:Ecorr} we note that the use of optimal orbitals compared to Hartree-Fock ones allows to gain additional $340$ keV. Although the effect is small, the variational aspect of the orbital equation is indeed found on these results. 
With the second truncation scheme, the energies have now decreased by more than $50$ MeV compared to the first scheme. Again, this overbinding is not surprising since the D1S Gogny interaction has not been fitted to reproduce binding energies at the MPMH configuration mixing level.
Beyond the missing two-body center-of-mass corrections, that are expected to unbind the nucleus by a few MeV, the huge total energy obtained when enlarging the valence space could be the result of different factors. First, we note that the diagonalization of the many-body matrix $\mathpzc{H}[\rho_{HF},\sigma_{HF}=0] = H[\rho_{HF}]+\mathpzc{R}[\rho_{HF},\sigma_{HF}=0]$ using the pure Hartree-Fock density as parameter in the Gogny interaction leads to a binding energy $E=-139.61$ MeV. Using the density of the correlated state thus brings additional $\sim 15$ MeV. The remaining overbinding could have two origins: 1) the density-dependent and spin-orbit terms being zero-ranged, they may lead to such pathological behaviors when the valence space is large and high relative momenta enter, 2) the T=0 residual part of the Gogny interaction is still not fully under control. Considering only relative energies, we note however the reasonable gain of $\sim 770$ keV when the full self-consistent process is applied. \\ 
At present, the uncertainties coming from the use of the D1S interaction prevent us from concluding on the efficiency of the method when using large valence spaces. Once a better constrained fully-finite range interaction will be developed and implemented, it would be very interesting to compare the results obtained in this approach with the ones produced by methods based on deformed mean-fields, such as the Generator Coordinate Method with angular momentum projection as described in e.g. \cite{RodriguezEgido}. Such a comparison would provide information on the ability of the MPMH approach to describe deformed nuclei.

\subsubsection{Wave function}
We show in Table~\ref{t:wf_trunc1} the effect of the orbital transformation on the composition of the ground state function, for the truncation scheme 1.

\begin{table}[h]
\centering
\begin{tabular}{cccc}
\hline     
Configuration       &  Eq.~(\ref{e:eq1_Gogny}) & Eqs.~(\ref{e:eq1_Gogny})$\&$(\ref{e:eq2_Gogny}) & Eqs.~(\ref{e:eq1_Gogny})$\&$(\ref{e:eq2_Gogny}) \\
                         &  (HF orbitals)         &  (HF orbitals)                               &  (SC orbitals)\\
\hline  
\hline
$0p$-$0h$                     &   53.95                &47.65                                         & 48.20         \\              
\hline
($2p$-$2h$)$_\pi$             &   8.94                 &9.74                                          &  9.87         \\
$0p_{3/2}\rightarrow 0p_{1/2}$&                        &                                              &               \\
\hline
($2p$-$2h$)$_\nu$             &   8.90                 &9.76                                          &  9.85         \\
$0p_{3/2}\rightarrow 0p_{1/2}$&                        &                                              &               \\
\hline
($2p$-$2h$)$_{\pi\nu}$        &  19.05                 &29.26                                         &  20.84        \\   
$0p_{3/2}\rightarrow 0p_{1/2}$&                        &                                              &               \\
\hline
others                        &   9.16                 &2.46                                          &  11.24        \\
within $0p$-shell             &                        &                                              &               \\
\hline
\end{tabular}
\caption{Truncation scheme 1. Weight of the main configurations in the correlated ground state without and with orbital optimization. HF stands for "Hartree-Fock", while SC stands for "Self-Consistent".}
\label{t:wf_trunc1}
\end{table}

\noindent With frozen Hartree-Fock orbitals (first column), the $0p$-$0h$ configuration embodies $\sim 54 \%$ of the total wave function. The correlated ground state is already importantly fragmented at this level, since spherical orbitals are used. This low component reflects the importance of correlations associated to deformation. After optimizing the orbitals, one can analyze the composition of the wave function in terms of optimal self-consistent orbitals, or in terms of the initial pure Hartree-Fock orbitals (by calculating overlaps of Slater determinants). We show the corresponding results on the third and second columns respectively. In terms of HF orbitals, we note that the ground state is further fragmented in the sense that the $0p$-$0h$ component is decreased by $\sim 6 \%$. The $2p$-$2h$ excitations of proton-neutron type within the $0p$-shell are enhanced by more than $10 \%$. This again shows the importance of proton-neutron correlations for the description of deformed nuclei, probably increased by the fact that the nucleus has $N=Z$. Looking now at the ground state composition in terms of self-consistent orbitals, we note that the $0p$-$0h$ component is slightly more important than the HF ground state. Although the effect is very small in this case, it shows that the new self-consistent reference state is "better" than the initial one, in the sense that it embodies more weight of the correlated wave function and thus incorporates more physical content. \\ \\

\noindent Similarly, we show in Table~\ref{t:wf_trunc2} the weight of the $0p$-$0h$ configuration before and after orbital renormalization, when using the truncation scheme 2. 

\begin{table}[h]
\centering
\begin{tabular}{cccc}
\hline     
    &  Eq.~(\ref{e:eq1_Gogny})  &  Eqs.~(\ref{e:eq1_Gogny}) $\&$ (\ref{e:eq2_Gogny})   &   Eqs.~(\ref{e:eq1_Gogny}) $\&$ (\ref{e:eq2_Gogny}) \\
    &  (HF orbitals)            &  (HF orbitals)                                       &  (SC orbitals) \\
\hline
\hline
$0p$-$0h$   &   21.46                   &20.39                                                 & 22.33 \\         
\hline
\end{tabular}
\caption{Truncation scheme 2. Weight of the $0p$-$0h$ configuration in the wave function without and with orbital optimization.}
\label{t:wf_trunc2}
\end{table}

\noindent The wave function is much more fragmented than in the previous case before optimizing the single-particle states. Again, the orbital transformation slightly increases this fragmentation (second column), and the optimized self-consistent reference state has a higher component than the initial HF state (third column). Let us note that the transformation of orbitals applied with the scheme 1 brings the wave function towards the fragmentation obtained with the scheme 2 -- which should be a better approximation to the "exact" solution (given the interaction). However this decrease of $6\%$, although non negligible, is not sufficient to reach the solution of the scheme 2. As already discussed in section \ref{section1}, we remind that the orbital variation only allows to optimize the one-body quantities in order to "absorb" part of the correlation content of the system. Changing the orbitals, however, does not introduce new correlations and thus will never fully make up for the truncation of the many-body space. This was also illustrated by Eq.~(\ref{e:PQcoupling}) which showed that the orbital equation allows to connect the subspaces $\mathcal{P}$ and $\mathcal{Q}$ via $H_{\mathcal{P}\mathcal{Q}}$, but does not account for $H_{\mathcal{Q}\mathcal{Q}}$.\\ \\

Finally let us remember what has been discussed in section \ref{section1}. The initial $\mathcal{P}$-space, denoted by $\mathcal{P}^{(i)}$, contains the many-body configurations $\ket{\phi_\beta^{(i)}}$ that are built on initial Hartree-Fock single-particle states and that are selected by the chosen truncation scheme. Similarly the final $\mathcal{P}$-space, $\mathcal{P}^{(f)}$, is the space of selected configurations $\ket{\phi_\alpha^{(f)}}$ built on optimal self-consistent orbitals. The idea here is to evaluate how much of the initially ignored $\mathcal{Q}$-space, $\mathcal{Q}^{(i)}$, has been incorporated into the final state $\ket{\Psi^{(f)}}$ via the optimization of orbitals. 
That is, writing 
\begin{eqnarray}
 \ket{\Psi^{(f)}} &=& \sum_{\alpha \in \mathcal{P}^{(f)}} A_\alpha^{(f)} \ket{\phi_\alpha^{(f)}} \nonumber \\
                  &=& \sum_{\beta \in \mathcal{P}^{(i)}} A_\beta^{(i)} \ket{\phi_\beta^{(i)}} + \sum_{\beta \in \mathcal{Q}^{(i)}} A_\beta^{(i)} \ket{\phi_\beta^{(i)}} \; ,
\end{eqnarray}
we want to evaluate the values of 
\begin{eqnarray}
 W_{\mathcal{P}^{(i)}} &\equiv& \sum_{\beta \in \mathcal{P}^{(i)}} |A_\beta^{(i)}|^2 \; , \hbox{   and} \nonumber \\
 W_{\mathcal{Q}^{(i)}} &\equiv& \sum_{\beta \in \mathcal{Q}^{(i)}} |A_\beta^{(i)}|^2  \; ,
\end{eqnarray}
which represents the weights of $\mathcal{P}^{(i)}$ and $\mathcal{Q}^{(i)}$ in the final wave function, respectively.
This is done by calculating the overlaps of configurations $\ket{\phi_\beta^{(i)}}$ with configurations $\ket{\phi_\alpha^{(f)}}$, since
\begin{equation}
 A_\beta^{(i)}= \braket{\phi_\beta^{(i)}|\Psi^{(f)}} = \sum_{\alpha\in \mathcal{P}^{(f)}} A_\alpha^{(f)}  \braket{\phi_\beta^{(i)}|\phi_\alpha^{(f)}} \; .
\end{equation}

\noindent We perform the calculation for $W_{\mathcal{P}^{(i)}}$ and deduce $W_{\mathcal{Q}^{(i)}}$ from the normalization condition $W_{\mathcal{P}^{(i)}}+W_{\mathcal{Q}^{(i)}}=1$. The results obtained with the truncation scheme 1 are displayed on the first line of Table~\ref{t:PQweight}. 

\begin{table}[h]
 \centering
 \begin{tabular}{ccc}  
 \hline
 valence space  \hspace{0.5cm}       &  $W_{\mathcal{P}^{(i)}}$ \hspace{0.5cm}  &  $W_{\mathcal{Q}^{(i)}}$ \\
\hline 
\hline
0p-shell        \hspace{0.5cm}       &  98.87 $\%$  \hspace{0.5cm}              &  {\bf 1.13} $\%$         \\
\hline      
0s-0p           \hspace{0.5cm}       &  97.42 $\%$  \hspace{0.5cm}              &  {\bf 2.58} $\%$         \\
\hline
0s-0p-1s0d      \hspace{0.5cm}       & 95.87 $\%$   \hspace{0.5cm}              &  {\bf 4.13} $\%$         \\
\hline
0s-0p-1s0d-0f1p \hspace{0.5cm}       &  96.93 $\%$  \hspace{0.5cm}              &   {\bf 3.07} $\%$        \\
\hline
\end{tabular}
\caption{Weight of the initial $\mathcal{Q}$ space introduced in the final wave function via the optimization of orbitals, according to the size of the valence space.}
\label{t:PQweight}
\end{table}

In this case only $\sim 1 \%$ of the initial $Q$ space is incorporated into the final correlated wave function. Again this may be due to the fact that only the $0p$-shell has been explicitly introduced in the configuration mixing and thus, because of the symmetry preservation, only the $p$ shells are directly impacted by the source term of the orbital equation. In other words, not enough correlations have been explicitly introduced at the beginning to have a noticeable response from the orbital equation. Results obtained for larger valence spaces are displayed on the next lines of Table~\ref{t:PQweight}. We note that the value of $W_{\mathcal{Q}^{(i)}}$ starts by increasing when adding the $0s$ shell to the active space, and again when adding the $sd$ shell. Introducing more active shells, and thus different angular momentum and parities in the mixing, seems to increase the effect of the orbital equation. However, when one continues to enlarge the active space by adding the $fp$-shell, the value of $W_{\mathcal{Q}^{(i)}}$ starts decreasing again. At this point the initial wave function is already close to the "exact" solution (for a given interaction and given size of the single particle basis). It becomes then less and less necessary to optimize the single-particle states. Therefore, it seems that there exists a size of valence space for which the orbital equation has a maximum effect.
Finally let us remind that the D1S interaction used here contains a density-dependence which already implicitly accounts for part of the space $\mathcal{Q}$. For a more conclusive study, it would therefore be very informative to perform the same analyses using an other type of interaction. Perhaps the results would be more striking in this case.

\subsection{Description of the first 2$^{+}$ excited state} \label{excited}

\begin{figure}[h]
\begin{center}
\includegraphics[width=0.7\columnwidth]{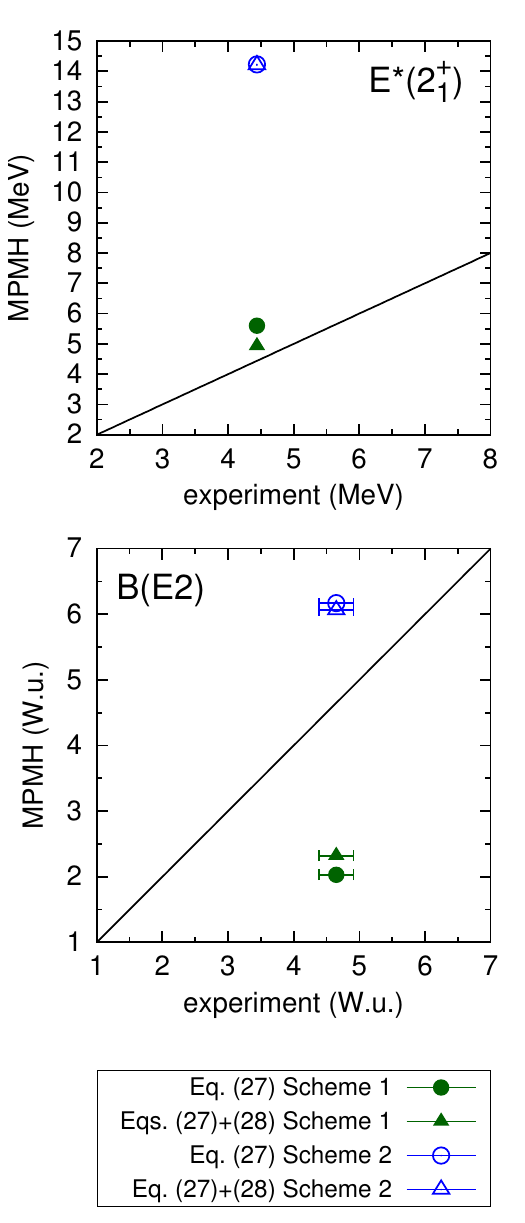}
\caption{(Color online) Excitation energy $E^*(2^+_1)$ in MeV (top) and transition probability $B(E2:2_1^+ \rightarrow 0_1^+)$ in Weisskopf units (bottom) for the first excited $2^+$ state.}
\label{f:E-BE2}
\end{center}
\end{figure}

We end this study with an analysis of the first $2^{+}$ excited state. Numerically, excited states are obtained in the following way: First, the complete convergence procedure described in section \ref{section2} is applied to the ground state of the nucleus. Then, when convergence is reached, we extract several eigenvalues of the many-body matrix $\mathpzc{H}[\rho,\sigma]$, corresponding to different many-body states. In other words, the orbitals are optimized according to the correlations of the ground state and also used to expand the excited states. Of course in principle one should solve both variational equations, (\ref{e:eq1_Gogny}) and (\ref{e:eq2_Gogny}), for each state and thus obtain a different set of orbitals for each of them. However, the approximation that we use is usually justified for the description of low-lying states.
\\
We show in Fig.~\ref{f:E-BE2} the theoretical values for the excitation energy $E^*(2^+_1)=E(2^+_1)-E(0^+_1)$ and the transition probability  $B(E2:2_1^+ \rightarrow 0_1^+)$ obtained with both truncation schemes. The results obtained with pure HF single-particle states are represented by circles while the ones obtained with self-consistent orbitals are represented by triangles. These values are compared to the experimental data \cite{nndc}.
\\
The truncation scheme 1 yields a good description of the energy, which is improved by $\sim 670$ keV when using self-consistent orbitals. The transition probability $B(E2)$ is however clearly underestimated. When using pure Hartree-Fock orbitals, the theoretical value differs from the experimental one by a factor $\sim 2.3$. It is well known that in order to reproduce the quadrupole collectivity, one usually needs to include explicit $2\hbar\Omega$ excitations in the valence space. Such a discrepancy is therefore expected when the mixing is restricted to the $0p$-shell ($0\hbar\Omega$ space) and no effective charge is used. The $B(E2)$ obtained with self-consistent single-particle states is slightly improved and differs from experiment by a factor $\sim 2$. This small improvement corresponds to the $1.13\%$ of the space $\mathcal{Q}$ that has been included in the wave function, as shown in Table~\ref{t:PQweight}.
\\
Enlarging the valence space allows to increase the collectivity and improves the description of the $B(E2)$. However the effect appears to be too important, so that the theoretical value now sits above the experimental one. Concerning the excitation energy, the good description that was obtained with the scheme 1 is not reproduced with the scheme 2 which largely overestimates the experimental $E^*(2^+_1)$ by $\sim 10$ MeV. This again shows that the correlations added to the ground state when enlarging the active space are incorrect, and are likely related to the D1S interaction we are using that already empirically incorporates correlations in its parameters. The zero-range spin-orbit and density-dependent terms are also to investigate. These might explain why the couplings between high-energy Slater determinants do not only bring collectivity, as it should be the case due to their chaotic behavior \cite{Pillet2}. As already noted in the previous section, when the full single-particle space is active, the transformation of orbitals, although acting here in the right way, has only very little impact on the results.

\section{Summary, conclusion and perspectives} \label{conclu}
In this work, we have fully applied for the first time the self-consistent multiparticle-multihole configuration mixing approach 
to the description of a nucleus. Based on a variational principle determining both the expansion coefficients of the wave function and the single-particle orbitals, this method establishes a natural bridge between Configuration-Interaction techniques and Self-Consistent Mean-Field methods. In order to gain insight into the equations we have exposed and analyzed the formalism in a general context considering a three-body Hamiltonian. A first application was then performed using the two-body density-dependent D1S Gogny force. We chose $^{12}$C as an example to test and compare the numerical algorithm using different truncation schemes of the wave function. Starting from a spherical Hartree-Fock solution, convergence has been successfully reached in a reasonable number of iterations, with convergence criteria of $1.0 \times 10^{-4}$ set on the one-body density matrix. Particular attention has been paid to the effect induced by the orbital equation on the description of different quantities. In particular, introducing the effect of correlations into the mean-field led to a global compression of single-particle spectra and gaps at the Fermi level. The effect was particularly strong when a $N\hbar\Omega$ valence space was used. Concerning ground-state properties, optimizing the single-particle states globally increased the fragmentation of the wave function. The comparison of different valence spaces revealed that there may exist a truncation scheme for which the effect of the orbital equation toward a better solution is maximal. However this equation acting at the one-body level only, the induced effect usually remains small. The D1S force which already partly accounts for the rest of the many-body space may also reduce the impact of the orbital transformation. Finally the calculation of binding and excitation energies led to unrealistic results when using a large valence space. This confirms that the D1S Gogny force is not prepared for this type of truncation scheme, and a better constrained fully finite range interaction is necessary in order to validate the method. Work in this direction is in progress \cite{D2,PilletPena}.
\\ 
In a next paper, we will apply the MPMH method using the numerical algorithm described in this work, to perform a systematic study of $sd$-shell nuclei. Ground-state properties and spectroscopy will be exposed. The structure description provided by MPMH will also be used as input to calculate cross sections associated with inelastic scattering processes.

\begin{acknowledgments}
N.P. would like to thank D. Gogny for valuable discussions about the formalism of this approach. C.R. and N.P. are indebted to V. Zelevinsky for his interest in the subject and for fruitful discussions, and also thank the National Superconducting Cyclotron Laboratory at Michigan State University for hospitality.
\end{acknowledgments}


\appendix*

\section{The orbital equation from the Green's function formalism at equal times} 
\label{a1}
\subsection{Derivation of the orbital equation}
Let us first recall the definition of the Green's functions.
Let $a^{\dagger}_{i}(t)$ and $a_i(t)$ be the creation and destruction operators of a particle in state $i$ in the Heisenberg picture. They are related to the time-independent operators $a^{\dagger}_i$ and $a_i$ in the Schroedinger picture as (with 
$\hbar=1$),
\begin{numcases}{}
a^{\dagger}_{i}(t) = e^{iHt} a^{\dagger}_i e^{-iHt} \nonumber \\
a_{i}(t) = e^{iHt} a_i e^{-iHt} \; .
\end{numcases} 
The many-body Green's functions in the representation $i$ are defined as follows,
\begin{numcases}{}
\mathcal{G}_{ij}^{[1]} (t_1-t_2) = -i \braket{\Psi|\mathcal{T}\left(a_i(t_1)a_j^{\dagger}(t_2)\right)|\Psi} \nonumber \\
\mathcal{G}_{ij,kl}^{[2]} (t_1,t_2;t_3,t_4) = - \braket{\Psi|\mathcal{T}\left(a_i(t_1)a_j(t_2)a^{\dagger}_l(t_4)a^{\dagger}_k(t_3)\right)|\Psi} \nonumber \\
\mathcal{G}_{ijk,lmn}^{[3]} (t_1,t_2,t_3;t_4,t_5,t_6) = \nonumber \\
\hspace{1.35cm}i \braket{\Psi|\mathcal{T}\left(a_i(t_1)a_j(t_2)a_k(t_3)a^{\dagger}_n(t_6)a^{\dagger}_m(t_5)a^{\dagger}_l(t_4)\right)|\Psi} \nonumber
\\
\hspace{1cm} \vdots \nonumber \\
\label{e:GFs}
\end{numcases}
where $\ket{\Psi}$ is in principle the exact ground-state of the A-particle system, and $\mathcal{T}()$ is the time-ordering operator which brings the
operators taken at latter times on the left of operators taken at earlier times and affects the results by the sign of the corresponding permutation.
\\
\\
The equation of motion for the one-body propagator can be obtained from the equation of motion for the Heisenberg annihilation operator $a_i(t)$. Considering a three-body Hamiltonian $\hat H = \hat{K}+\hat{V}^{2N}+\hat{V}^{3N}$, we have
\begin{eqnarray}
 i\frac{\partial}{\partial t}a_{i}(t) 
 &=& [a_{i}(t),\hat{H}] = e^{i\hat{H}t}[a_{i},\hat{H}]e^{i\hat{H}t} \nonumber \\
 &=& \sum_{j k} K_{j k}  e^{i\hat{H}t}[a_{i},a_{j}^{\dagger}a_{k}]e^{i\hat{H}t} \nonumber \\
 && + \frac{1}{4} \sum_{j k, l m} \widetilde{V}^{2N}_{j k ,l m}  e^{i\hat{H}t} [a_{i},a_{j}^{\dagger}a_{k}^{\dagger}a_{m}a_{l}]e^{i\hat{H}t} \nonumber \\
 && + \frac{1}{36} \sum_{sqklmn}  \widetilde{V}^{3N}_{sqk,lmn}  e^{i\hat{H}t}
      [a_{i},a_{s}^{\dagger}a_{q}^{\dagger}a_{k}^{\dagger}a_{n}a_{m}a_{l}]e^{i\hat{H}t} \nonumber \\
 &=& \sum_{k} K_{i k} a_{k}(t) + \frac{1}{2} \sum_{k l m} \widetilde{V}^{2N}_{i k, l m} a_{k}^{\dagger}(t)a_{m}(t)a_{l}(t) \nonumber \\
 &&  + \frac{1}{12} \sum_{qklmn} \widetilde{V}^{3N}_{iqk ,lmn} a_{q}^{\dagger}(t)a_{k}^{\dagger}(t)a_{n}(t)a_{m}(t)a_{l}(t) \; . \nonumber \\
\label{e:eom_3body}  
\end{eqnarray} 

\noindent Multiplying (\ref{e:eom_3body}) by $a_{j}^{\dagger}(t')$ on the right, taking the $\mathcal{T}$-product and the expectation value in  $\ket{\Psi}$ of the corresponding expression, and using the fact that the $\mathcal{T}$-product of operators is a distribution we finally obtain
\begin{eqnarray}
 && \sum_k \left( i\delta_{ik} \frac{\partial}{\partial t}  - K_{ik} \right) \mathcal{G}_{k j} (t-t') \nonumber \\
 &&  =   \delta(t-t') \delta_{i j} + \frac{i}{2} \sum_{k l m} \widetilde{V}^{2N}_{i k, l m} \mathcal{G}^{[2]}_{m l ,j k} (t , t ; t' , t^+) \nonumber \\
 && \hspace{0.2cm}- \frac{1}{12} \sum_{qklmn} \widetilde{V}^{3N}_{iqk,lmn} \mathcal{G}^{[3]}_{nml,jkq} (t , t ,t ; t' , t^+, t^+) \; . 
\label{e:EoM_GF1a}
\end{eqnarray} 
This is the equation of motion expressing the one-body propagator $\mathcal{G}^{[1]}$ in term of $\mathcal{G}^{[2]}$ and $\mathcal{G}^{[3]}$, first step of the famous Martin-Schwinger hierarchy of equations \cite{MartinSchwinger}.
\\ 
\\
\noindent Similarly one can repeat the same steps starting from the equation of motion for $a_j^{\dagger}(t')$. This leads to the following equivalent equation
\begin{eqnarray}
 && \sum_k G_{ik} (t-t') \left( i \frac{\overleftarrow{\partial}}{\partial t'} \delta_{kj} + K_{kj} \right)  \nonumber \\  
 &&=   - \delta(t-t') \delta_{i j} + \frac{i}{2} \sum_{k l m} \widetilde{V}^{2N}_{k l, m j} \mathcal{G}^{[2]}_{im,lk} (t , t'^- ; t' , t') \nonumber \\
 &&+  \frac{1}{12} \sum_{k l m n p} \widetilde{V}^{3N}_{klm,nqj} \mathcal{G}^{[3]}_{iqn,mlk} (t , t'^-, t'^- ; t' , t', t') \; . 
\label{e:EoM_GF1b}
\end{eqnarray}

\noindent Adding Eq.~(\ref{e:EoM_GF1a}) to (\ref{e:EoM_GF1b}) eliminates the time derivatives and we get
\begin{eqnarray}
&&  \sum_{k} \left(- K_{i k} \mathcal{G}^{[1]}_{k j} (t-t') + \mathcal{G}^{[1]}_{i k} (t-t') T_{k j} \right) \nonumber \\ 
&&= \frac{i}{2} \sum_{k l m} \widetilde{V}^{2N}_{i k, l m} \mathcal{G}^{[2]}_{m l, j k} (t , t ; t' , t^+) \nonumber \\
&&+ \frac{i}{2} \sum_{k l m} \widetilde{V}^{2N}_{k l, m j} \mathcal{G}^{[2]}_{i m ,l k} (t , t'^- ; t' , t')  \nonumber \\ 
&&- \frac{1}{12} \sum_{qklm} \widetilde{V}^{3N}_{iqk,lmn} \mathcal{G}^{[3]}_{nml,jkq} (t , t ,t ; t' , t^+, t^+) \nonumber \\
&&+ \frac{1}{12} \sum_{k l m n q} \widetilde{V}^{3N}_{klm,nqj} \mathcal{G}^{[3]}_{iqn,mlk} (t , t'^-, t'^- ; t' , t', t') \; . 
\label{e:eq2-time}
\end{eqnarray}

\noindent We now want to take the equal-time limit $t' \rightarrow t^+$ of Eq.~(\ref{e:eq2-time}). It is straightforward to see that, in this limit, the N-body propagator is proportional to the N-body density. In particular we have for the one-body propagator
\begin{eqnarray}
\lim_{t' \rightarrow t^+} \mathcal{G}^{[1]}_{k j} (t-t') &=& -i \braket{\Psi|T\left( a_{k}(t) a_{j}^{\dagger}(t^+) \right) |\Psi } \nonumber \\
                                                         &=& +i \rho_{k j} \; ,
\label{e:G1_eq_time}                                                
\end{eqnarray}
for the two-body propagator
\begin{eqnarray}
\lim_{t' \rightarrow t^+} \mathcal{G}^{[2]}_{m l, j k} (t , t ; t' , t^+) &=& - \braket{ \Psi| T \left( a_{m}(t)  a_{l}(t)  a_{k}^{\dagger}(t^+)  a_{j}^{\dagger}(t^+) \right)|\Psi } \nonumber \\
                                                                          &=& + \braket{ \Psi| a_{j}^{\dagger} a_{k}^{\dagger} a_{m} a_{l}    |\Psi } \nonumber \\
                                                                          &=& \rho_{l j} \rho_{m k} - \rho_{l k} \rho_{m j} + \sigma_{jl , km} \; ,
\label{e:G2_eq_time}
\end{eqnarray}
and for the three-body propagator
\begin{eqnarray}
  &&\lim_{t' \rightarrow t^+} \mathcal{G}^{[3]}_{nml,jkq} (t,t,t; t',t^+,t^+) \nonumber \\
  &&= +i \braket{ \Psi| T \left( a_{n}(t) a_{m}(t)  a_{l}(t) a_{q}^{\dagger}(t^+)  a_{k}^{\dagger}(t^+)  a_{j}^{\dagger}(t^+) \right)|\Psi } \nonumber \\
  &&= -i \braket{\Psi| a^\dagger_q a^\dagger_k a^\dagger_j a_n a_m a_l  |\Psi} \nonumber \\
  &&= -i \Bigl( \rho_{lq} \rho_{mk} \rho_{nj} - \rho_{lq} \rho_{mj} \rho_{nk} - \rho_{lk} \rho_{mq} \rho_{nj} \nonumber \\
  && + \rho_{lk} \rho_{mj} \rho_{nq} - \rho_{lj} \rho_{mk} \rho_{nq} + \rho_{lj} \rho_{mq} \rho_{nk} \nonumber \\
  && + \rho_{lq} \sigma_{km,jn} - \rho_{lk} \sigma_{qm,jn} - \rho_{lj} \sigma_{km,qn} \nonumber \\
  && + \rho_{mk} \sigma_{ql,jn} - \rho_{mq} \sigma_{kl,jn} - \rho_{mj} \sigma_{ql,kn} \nonumber \\
  && + \rho_{nj} \sigma_{ql,km} - \rho_{nq} \sigma_{jl,km} - \rho_{nk} \sigma_{ql,jm} \nonumber \\
  && + \chi_{ql,km,jn} \Bigr) \; ,
\label{e:G3_eq_time}
\end{eqnarray}
where $\sigma$ and $\chi$ are the two- and three-body correlation matrices. They correspond to the equal-time limit of the connected two- and three-body propagators respectively. Using relations (\ref{e:G1_eq_time})-(\ref{e:G3_eq_time}) to take the limit $t' \rightarrow t^+$ of Eq.~(\ref{e:eq2-time}) finally leads after calculus exactly to the orbital equation (\ref{e:eq2}):
\begin{equation}
\left[\hat{h}[\rho,\sigma],\hat{\rho}\right] =\hat{G}[\rho,\sigma,\chi] \nonumber \; ,
\end{equation} 
where $h[\rho,\sigma]$ and $G[\rho,\sigma,\chi]=F[\rho,\sigma,\chi]-F^\dagger[\rho,\sigma,\chi]$ are defined in Eq.~(\ref{e:mean_field}) and (\ref{e:G_3N}).

\subsection{Relation to the Dyson equation}
As it is well-known, the equation of motion (\ref{e:EoM_GF1a}) for the one-body propagator can equivalently be written in terms of the free propagator $\mathcal{G}^{[0]}$ satisfying 
\begin{equation}
\sum_l \left( i\delta_{il} \frac{\partial}{\partial t}  - K_{il} \right) \mathcal{G}^{[0]}_{l j} (t-t') =  \delta(t-t') \delta_{i j} \; ,
\end{equation} 
as
\begin{eqnarray}
 &&\mathcal{G}^{[1]}_{ij} (t-t')  = \mathcal{G}^{[0]}_{ij} (t-t')  \nonumber \\
 &&+ \sum_{ks} \int \D t_1 \int \D t_2 \mathcal{G}^{[0]}_{ik}(t-t_1) \Sigma_{ks}(t_1-t_2) \mathcal{G}^{[1]}_{sj}(t_2-t') \; ,  \nonumber \\
\label{e:Dyson_a}
\end{eqnarray}
where $\Sigma$ is the so-called "self-energy" which contains all the information about the many-body propagators, and is defined as
\begin{eqnarray}
&&\sum_{s} \int \D t_2 \Sigma_{is}(t_1-t_2) \mathcal{G}^{[1]}_{sj} (t_2-t') \nonumber \\
&&\hspace{0.5cm} =\frac{i}{2} \sum_{lmn} \widetilde{V}^{2N}_{il,mn} \mathcal{G}^{[2]}_{nm,jl}(t_1,t_1;t',t_1^+) \nonumber \\ 
&&\hspace{0.7cm} -\frac{1}{12} \sum_{qklmn} \widetilde{V}^{3N}_{iqk,lmn} \mathcal{G}^{[3]}_{nml,jkq} (t_1 , t_1 ,t_1 ; t' , t_1^+, t_1^+) \; .\nonumber \\
\label{e:self}
\end{eqnarray} 
Eq.~(\ref{e:Dyson_a}) is the famous Dyson equation.
\\
\\
The self-energy can always be split into a static part $\Sigma^{[0]}$ and a dynamical part $\Sigma^{dyn}$ as 
\begin{equation}
\Sigma(t-t') = \Sigma^{[0]} \delta(t-t') + \Sigma^{dyn} (t-t') \; .
\end{equation}
In Eq.~(\ref{e:self}), one can express the two-body propagator in term of its connected part $\mathcal{G}^{[2]C}$ as
\begin{eqnarray}
&&\mathcal{G}^{[2]}_{ij,kl}(t_1,t_2;t_3,t_4)\nonumber \\
&&=\mathcal{G}^{[1]}_{ik}(t_1-t_3)\mathcal{G}^{[1]}_{jl}(t_2-t_4)-\mathcal{G}^{[1]}_{il}(t_1-t_4)\mathcal{G}^{[1]}_{jk}(t_2-t_3) \nonumber \\
&&\hspace{0.8cm}+\mathcal{G}^{[2]C}_{ij,kl}(t_1,t_2;t_3,t_4) \; .
\end{eqnarray} 
Similarly the three-body propagator can be written in terms of antisymmetrized products of the type $\mathcal{G}^{[1]}\mathcal{G}^{[1]}\mathcal{G}^{[1]}$, $\mathcal{G}^{[1]}\mathcal{G}^{[2]C}$ and its connected part $\mathcal{G}^{[3]C}$. It is then easily shown that the static self-energy $\Sigma^{[0]}$ corresponds to the average potential defined in Eq.~(\ref{e:mean_field}):
\begin{equation}
\Sigma^{[0]} = \Gamma[\rho,\sigma]\equiv\Gamma^{2N}[\rho]+\Gamma^{3N}[\rho,\sigma] \; .
\end{equation}
$\;$\\
The dynamical self-energy is then given by
\begin{eqnarray}
 &&\Sigma_{ij}^{dyn}(t-t') = \nonumber \\
 &&-i \int dt_1 \sum_{klmn} \widetilde{V}^{2N}_{ki,lm} \mathcal{G}^{[2]C}_{ml,nk}(t,t;t_1,t^+) \mathcal{G}^{[1]-1}_{nj} (t_1-t') \nonumber \\
 &&-\frac{i}{2} \int dt_1  \sum_{qklmns} \widetilde{V}^{3N}_{iqk,lmn} \rho_{mk} \mathcal{G}^{[2]C}_{nl,jq}(t,t;t_1,t^+) \mathcal{G}^{[1]-1}_{nj} (t_1-t') \nonumber \\
 && - \frac{1}{12} \int dt_1  \sum_{qklmns} \widetilde{V}^{3N}_{iqk,lmn}  \mathcal{G}^{[3]C}_{nml,jkq}(t,t,t;t_1,t^+,t^+) \mathcal{G}^{[1]-1}_{nj} (t_1-t') \; ,\nonumber \\
\end{eqnarray} 
and can be related to the source term $G[\rho,\sigma,\chi]=F[\rho,\sigma,\chi]-F^\dagger[\rho,\sigma,\chi]$ of the orbital equation since
\begin{eqnarray}
 &&\lim_{t_2\rightarrow t_1^+}  \sum_s \int \D t' \Sigma^{dyn}_{is}(t_1-t') \mathcal{G}^{[1]}_{sj} (t'-t_2) \nonumber \\
 && = \frac{i}{2}\sum_{klm} \widetilde{V}^{2N}_{ki,lm} \sigma_{kl,jm} -\frac{i}{2} \sum_{qklmn} \widetilde{V}^{3N}_{iqk,lmn} \rho_{mk} \sigma_{jl,kn} \nonumber \\
 && \hspace{0.4cm}+ \frac{i}{12} \sum_{qklmn} \widetilde{V}^{3N}_{iqk,lmn} \chi_{ql,km,jn} \nonumber \\                                                                                                                                            && = i (F^{\dagger})_{ij} \; .
\end{eqnarray}

\end{document}